\documentclass[aps,prb,showpacs,twocolumn,floats,epsfig]{revtex4-2}
\usepackage{amssymb}
\usepackage{amsbsy}
\usepackage{amsmath}
\usepackage{epsfig}
\usepackage{color}
\usepackage{float}
\usepackage{xr-hyper}
\usepackage[colorlinks,linktocpage,bookmarks=false,citecolor=blue,linkcolor=red,urlcolor=blue]{hyperref}

\newcommand{\ing}{\includegraphics}

\newcommand{\beg}{\begin{gather}}
\newcommand{\eeg}{\end{gather}}
\newcommand{\beq}{\begin{equation}}
\newcommand{\eeq}{\end{equation}}
\newcommand{\bea}{\begin{eqnarray}}
\newcommand{\eea}{\end{eqnarray}}

\newcommand{\ket}[1]{\ensuremath{\left|#1\right\rangle}}

\newcommand\mean[1]{\ensuremath{\left\langle#1\right\rangle}}

\newcommand\lrp[1]{\left (#1\right)}
\newcommand\lrb[1]{\left[#1\right]}
\newcommand\lrc[1]{\left\{#1\right\}}

\newcommand{\lra}{\quad \Leftrightarrow \quad}

\newcommand{\be}{\begin{equation}}
\newcommand{\ee}{\end{equation}}
\def\ba{\begin{aligned}}
\def\ea{\end{aligned}}
\def\bes{\begin{subequations}}
\def\ees{\end{subequations}}
\def\bal{\begin{align}}
\def\eal{\end{align}}

\newcommand{\imk}[1]{{\color{black}#1}}
\newcommand{\imkR}[1]{{\color{black}#1}}

\usepackage{soul}
\usepackage{ulem} 
\usepackage{hyperref}
\graphicspath{{./figures/}}

\begin{document}
\title{Tuning the phase diagram of a Rosenzweig-Porter model with fractal disorder}
\author{Madhumita Sarkar }
\thanks{sarkar.madhumita770@gmail.com}
\affiliation{Jo{\v z}ef Stefan Institute, SI-1000, Ljubljana, Slovenia}
\author{Roopayan Ghosh}
\affiliation{Department of Physics and Astronomy, University College London, Gower Street, WC1E6BT, London}
\author{Ivan Khaymovich}
\affiliation{Nordita, Stockholm University and KTH Royal Institute of
Technology Hannes Alfv{\'e}ns v{\"a}g 12, SE-106 91 Stockholm, Sweden}
\affiliation{Institute for Physics of Microstructures, Russian Academy of Sciences, 603950 Nizhny Novgorod, GSP-105, Russia}
\begin{abstract}
Rosenzweig-Porter (RP) model has garnered much attention in the last decade, as it is a simple analytically tractable model showing both ergodic--nonergodic extended and Anderson localization transitions. Thus, it is a good toy model to understand the Hilbert-space structure of many body localization phenomenon. In our study, we present analytical evidence, supported by exact numerics, that demonstrates the controllable tuning of the phase diagram in the RP model by employing on-site potentials with a non-trivial fractal dimension instead of the conventional random disorder. We demonstrate that \imk{such disorder} extends the fractal phase and creates unusual dependence of fractal dimensions of the eigenfunctions. \imk{Furthermore, we study the fate of level statistics in such a system to understand how these changes are reflected in the eigenvalue statistics.}
\end{abstract}
\maketitle
\paragraph*{\textbf{Introduction:}}
Disorder-induced breakdown~\cite{Evers2008Anderson,Basko06,gornyi2005interacting,Abanin_RMP} of quantum ergodicity~\cite{Deutsch1991,Srednicki1994}, dubbed as many-body localization~(MBL), is a generic phenomenon in many-body~(MB) systems. 
Being \imk{a localized phase} in real space, MBL \imk{provides} only ergodicity breaking in its Hilbert counterpart~\cite{Luitz15,Mace_Laflorencie2019_XXZ,QIsing_2021}. 
This fact as well as the discovery of non ergodic extended phases in MB systems~\cite{BarLev2015absence,Griffiths2015anomalous,Luitz2016extended,Luitz2016anomalous,Khait2016spin,Znidaric2016Diffusive,BarLev2017transport,Bera2017density,agarwal2017rare,luitz2017ergodic,Lezama2019apparent,roy2018anomalous} as an intermediate regime between ergodic and localized phases~\cite{Evers2008Anderson,Floquet_MF,Sarkarnonintmfrac,Sarkarintmfrac,SenPhysRevB.107.035402,SayakPhysRevLett.123.025301,AbaninPhysRevB.96.104201} necessitated the search for analytically tractable toy models to understand \imk{this phenomenon}. One direction of this search \imk{is based on} the random-matrix ensembles that mimic the Hilbert-space properties of such MB systems in a controlled fashion. The Rosenzweig-Porter (RP) random matrix model~\cite{RP} \imk{provides} such an example which has been studied extensively in recent years as it allows almost a complete analytical understanding of the phase diagram~\cite{Kravtsov_NJP2015,Biroli_RP,Ossipov_EPL2016_H+V, Monthus, BogomolnyRP2018,Venturelli2023replica}, and a perturbative (exact in the thermodynamic limit) description~\cite{vonSoosten2017non} of the eigenspectrum for a wide range of parameter values.

The RP model is given by the Gaussian random-matrix ensemble of size $L$, where each element is random number obtained from a normal distribution, and the off-diagonal elements are rescaled by a factor of $L^{-\gamma/2}$,
\begin{equation}
 H_{mn}=h_n \delta_{mn}+ M_{mn}L^{-\gamma/2},
\end{equation}
where $\overline{h_n}=\overline{M_{mn}}=0$, $\overline{h_n^2}=\overline{M_{mn}^2}=1$. It has been shown that~\cite{Kravtsov_NJP2015,Biroli_RP}, with increasing $\gamma$ from $0$ to large values, this model, first, exhibits ergodicity, and then undergoes a transition to nonergodic extended (fractal) phase at $\gamma=1$. In the fractal phase the eigenfunction 
\imk{support sets contain} extensive number, but measure zero of all the lattice sites, scaling as $L^{D}$, where $D=2-\gamma$ denotes the second fractal dimension of the eigenfunction \imk{(the precise definition of $D$ is provided in the next section)}. As immediately apparent, $D=1$ corresponds to the ergodic phase. \imk{Furthermore, at} $\gamma=2$, $D$ goes linearly to $0$, marking the onset of the Anderson localized phase. \imk{Although,} this version of the model lacks \imk{the} genuine multifractality in its eigenfunctions, recent developments~\cite{Monthus,LN-RP_RRG,LN-RP_WE,BirTar_Levy-RP,LN-RP_dyn,Kutlin2023no_multifractal} show that some modified versions of this model may exhibit multifractality. Further studies~\cite{2022_nonHerm_RP} demonstrated the instability of the nonergodic extended phase in non-Hermitian version of the model. 
\begin{figure}[H]
\centering {\ing[width=0.75\linewidth]{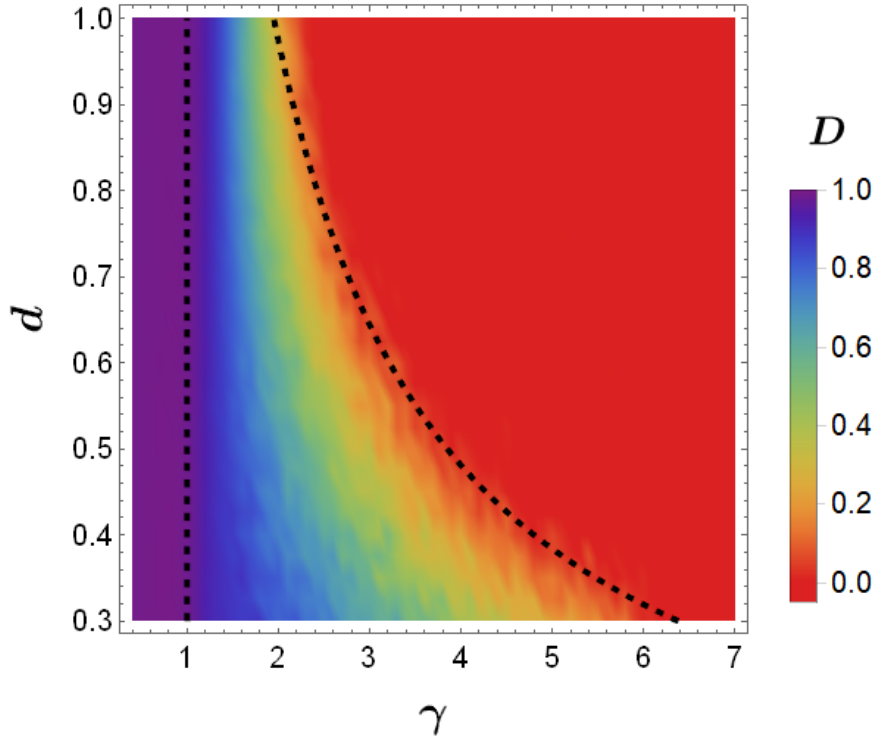}}
\caption{Fractal dimension \imk{$D$} vs $\gamma$ and the Hausdorff dimension $d$, showing good agreement with the analytical predictions of ergodic,  $\gamma_{ET} \simeq 1$ and Anderson $\gamma_{AT} \simeq 2/d$ transitions (black dotted lines), Eq.~\eqref{eq:Gamma_res}. 
Numerically, \imk{in Figs.~\ref{fig:fig1}-\ref{fig:d2mfrac}} $D$ has been obtained by fitting the inverse participation ratio (IPR)\imk{, Eq.~\eqref{eq:IPR},} of the eigenvectors for system sizes $L=2^p$, $7 \leq p \leq 12$.}
\label{fig:fig1}
\end{figure}
Unlike the latter case, in this work we show how to \imk{extend} the range and stability of the fractal phase of the RP model \imk{by employing a `fractal' on-site disorder}. \imk{We also include} the possibility of obtaining \imk{the} \imk{nonlinear dependence of fractal dimension on the parameter, $\gamma$.}

\paragraph*{\textbf{Summary of Results:}}
In our study, we selectively employ a random normal distribution solely for the off-diagonal elements of $H$, while the diagonal elements ($H_{mm}$) are sourced from a `fractal' disorder distribution with a Hausdorff dimension of $d$. This implies that there are typically $L^{1-d \times b}$ 
random diagonal elements, present in an energy window of width $L^{-b}$, if the total bandwidth is taken to be $O (1)$. One of the well-known examples of such a distribution is the Cantor set with a Hausdorff dimension~\cite{Kravtsov2023Cantor} $d=\ln2/\ln3$.

The main result of our work is shown in Fig.~\ref{fig:fig1}, where we represent the extended phase diagram of the RP model in terms of the second fractal dimension obtained from numerical fits of the generalized IPR,
\begin{equation}\label{eq:IPR}
 IPR^{(q)}_j=\sum_i^L|\langle i|\chi_j\rangle|^{2q},
\end{equation}
\imk{for $q=2$}; where $\ket{i}$ denote computational basis states and $\ket{\chi_j}$ denotes an eigenvector with index $j$. Then $IPR^{(2)}\sim L^{-D_2}$, \imk{where $D_2$ is the second fractal dimension}, one obtains $D_2$ from averages over numerical fits from a band of eigenvectors. $D_2$ can then be used to distinguish between \imk{the} ergodic ($=1$), nonergodic extended, i.e. fractal ($0<D_2<1$) and Anderson localized phase ($=0$). \imk{We refer} to $D_2$ as $D$ since the higher moments show the same value i.e. $D_q=D_2, q>1/2$, thus indicating that the eigenfunctions are fractal and not multifractal. In Fig.~\ref{fig:fig1}, we plot the \imk{$\gamma$- and $d$-dependence of $D$}. We note the following from the plot,
\begin{enumerate}
 \item $d=1$ represents the case for the generic RP model and the ergodic-fractal transition occurs at $\gamma=\gamma_{ET}=1$ and the Anderson transition occurs at $\gamma=\gamma_{AT}=2$ replicating the known results~\cite{Kravtsov_NJP2015}.
 \item As \imk{the} Hausdorff dimension of the diagonal elements decreases, \imk{$\gamma_{ET}$ is intact}, but $\gamma_{AT}$ monotonically increases, extending the fractal phase in $\gamma$.
 \item Both the transitions can be very well approximated by perturbative analytical expressions, which becomes exact in the thermodynamic limit, denoted by black dashed lines in the plot.
\end{enumerate}

It is also worth mentioning that, for $d>1$, the phase diagram shows similar behaviour as for $d=1$. This is because beyond the physical dimension of the diagonal disorder (1D in our Hermitian case), any increase in fractal dimension cannot have an effect~\footnote{Unlike the non-Hermitian case, where $1<d<2$ does change the phase diagram~\cite{2022_nonHerm_RP}.}.

In what follows, we, first, analytically \imk{calculate} the fractal dimension for eigenfunctions of the \imk{fractal} RP model with changing $\gamma$ and $d$. Then we compare the obtained expressions with exact numerics performed for (i)~the commonly studied Cantor set fractal distribution, and then (ii)~for a distribution with arbitrary Hausdorff dimensions $d$\imk{, suggested in~\cite{Kravtsov2023Cantor}. }
\imk{For completeness, we also discuss the level spacing statistics in such a model, and in the supplementary material~\cite{SM} discuss the time-dependent survival probability of a wave packet, initially localized at a single site.}

\paragraph*{\textbf{Analytical 
phase-diagram calculations:}} As mentioned before, we consider the $h_n$s to be distributed in a fractal (and later multifractal) manner~\footnote{Note that the strength of the diagonal elements is obtained from a fractal distribution, and not their spatial spread}. This implies that the $h_n$'s are distributed such that, the number ($\#$) of $h_n$'s in a given energy interval 
$|E-h_n| \in [L^{-b -
{d}b}, L^{-b}]$, parameterized by $b$ ($
{d}b\lesssim 1/\ln L$) 
vary as 
\begin{equation}
 \#\lrc{|E-h_n|\in \lrb{L^{-b-
 {d}b},L^{-b}}} \equiv L^{1-f (b)}
 {d}b
\label{eq:f (b)_def} ,
\end{equation}
with a certain $f (b)\imkR{\leq 1} $, characterizing the above fractal. We also assume that the overall bandwidth of the $h_n$ is $\sim O (1) = L^0$. Thus, $f (0)=0$. For any generic fractal with the Hausdorff dimension $d$ we will have, 
\be\label{eq:f (b)_Cantor}
f (b) = d \cdot b \ ,
\ee
For the special case of the Cantor set, $d=\ln 2/\ln 3$.
\imk{Note that, in general, $f(b)$ can depend on $E$, but for the case of the Cantor set $E$-dependence arises only in $1/\ln L$ corrections to $f(b)$ beyond the saddle-point expression~\eqref{eq:f (b)_def}.}
In contrast, in the case of uniform disorder distribution, the number of $h_n$'s is proportional to the width of the energy interval i.e., $f (b)=b$. Thus the usual Hermitian case~\cite{Kravtsov_NJP2015} corresponds to $d=1$, while the non-Hermitian complex one~\cite{2022_nonHerm_RP} gives $d=2$.

\imkR{The above saddle-point consideration in Eq.~\eqref{eq:f (b)_def} is valid as soon as the number $L^{1-f(b)}\gg 1$ is large. As we will see below, this corresponds to delocalized phases, where all the energy intervals are much larger than the typical level spacing, $\delta_{typ}$, i.e. the energy interval where one typically finds a single energy level.
Indeed,} the typical level spacing of the disorder, $\delta_{typ}$ is given by,
\begin{align}
 \#&=L^{1-f (b_{typ})}=1 \lra f (b_{typ}) =1 \nonumber \\ 
 &\lra b_{typ}=1/d \lra \delta_{typ}=L^{-b_{typ}}=L^{-1/d}
\end{align}
In this work, we focus only on real entries and, thus, work in the scenario $0<d<1$.
\imk{The generalization to the non-H}ermitian matrices to cover $0\leq d\leq 2$ is straightforward. In what follows, we provide a short description of computation of the fractal dimension of a typical eigenstate of this model and thus compute $\gamma_{AT}$ and $\gamma_{ET}$.

Using the \imk{standard} cavity Green's function method, we can find a self-consistency equation for the level broadening (the imaginary part of the self energy) $\Gamma_m$ 
as \imk{(see~\cite{SM}} 
and~\cite{Biroli_RP,Monthus,BogomolnyRP2018}),
\begin{equation}
 \bar{\Gamma}=\frac{1}{L}\sum_n \Gamma_n=\sum_n\frac{L^{-\gamma} ( \bar{\Gamma}-\eta)}{ (E-h_n)^2+ (\bar{\Gamma}-\eta)^2}
 \label{eq:selfconsistency1}
\end{equation}
where 
$E$ is the eigenenergy of the corresponding eigenvector \imk{and $\eta$ is a small regularizer. The parameterization $\bar{\Gamma}=L^{-a}$ in the limit $\eta \rightarrow 0$ gives the following result from Eq.~\eqref{eq:selfconsistency1} within the saddle-point approximation}, \imk{(see~\cite{SM})} 
\begin{align}
 1&\sim L^{1-\gamma +2a-f (a)} \lra \gamma = 1+2a-f (a) ,
\label{eq:self_consist_eq2}
\end{align}
This determines $\Gamma\sim L^{-a}$ via the parameter $\gamma$ \imkR{and works for $\Gamma\gg \delta_{typ}$}.  

\imk{The corresponding fractal dimension $D_q\equiv D$ is determined} via the number of levels located in the interval $\Gamma\sim L^{-a}$. \imk{This number} is related to the fractal dimension as $L^D$. From Eq.~\eqref{eq:f (b)_def}, we know that this is given by $L^{1-f (a)}$. Thus,
\be\label{eq:D_q}
D = 1-f (a) \ .
\ee
This definition of $D$ is the fractality in the ``space'' of $h_n$, but for the RP-like fractal phases it is equal to the spatial fractal dimension due to the Lorenzian structure of the eigenstates~\cite{Monthus,BogomolnyRP2018,RP_R(t)_2018,Buijsman2022circular,2022_nonHerm_RP}:
\be\label{eq:psi_Lorenz}
\mean{|\psi_{E} (n)|^2}_{H_{m\ne n}} \sim \frac{1}{ (E-h_n)^2 + \Gamma^2} \ .
\ee
As by fixing either $E$ or $h_n$, one has the Lorenzian, the fractality over the energy $E$ and over the ``space'' $h_n$ is equivalent to each other. In space $n$, the above Lorenzian forms a fractal miniband~\cite{Kravtsov2023Cantor} of the width $\Gamma$, with the underlying fractal structure $h_n$, living in that miniband, $|h_n - E|\lesssim \Gamma$.
 
\imk{In the fractal case of Eq.~\eqref{eq:f (b)_Cantor}, we obtain for $\gamma>1$ using Eqs.~\eqref{eq:self_consist_eq2}~-~\eqref{eq:D_q},}
\begin{align}\label{eq:Gamma_res}
{D = \max\lrp{1-d\frac{\gamma-1}{2-d},0}} \ , \quad \Gamma \sim L^{-\frac{\gamma-1}{2-d}} \ .
\end{align}
\imk{The Anderson transition point corresponds to} $\Gamma\simeq \delta_{typ}$, i.e., $D=0$, since in the localized phase the number of energy levels \imk{within the Lorenzian} bandwidth becomes an intensive quantity. Hence, $a = b_{typ}=1/d$ and
\begin{equation}
 \gamma_{AT} = 2/d \ .
 \label{eq:Andersontransition}
\end{equation} 
The ergodic transition occurs at $\Gamma\sim O(1)$, $\gamma_{ET} = 1$. Note that both $\gamma_{ET}$ and $\gamma_{AT}$ are continuous transitions, unlike the fat-tailed distributed RP models~\cite{LN-RP_RRG,LN-RP_WE,BirTar_Levy-RP}. 

\paragraph*{\textbf{Cantor Set diagonal elements:}} The first example we consider is when the diagonal elements are represented by Cantor set, $\mathcal{C}$. Cantor set is a set of points lying in a line segment normalized to the interval $[0,1]$, obtained by removing the middle third of the continuous line segments in a recursive manner. The set generated by the first few iterations of this are,
\begin{align}
 \mathcal{C}_0&=\lrb{0,1} \nonumber \\
 \mathcal{C}_1&=\lrb{0,\frac{1}{3}} \cup \lrb{\frac{2}{3},1}\nonumber \\
 \mathcal{C}_2&=\lrb{0,\frac{1}{9}} \cup \lrb{\frac{2}{9},\frac{1}{3}} \cup\lrb{\frac{2}{3},\frac{7}{9}} \cup\lrb{\frac{8}{9},1}\\
 \hdots \nonumber
\end{align}
We generate the diagonal elements by choosing the boundary value of each subset at the $n=\log_2 L$ iteration. The self similar nature of the Cantor set is evident from the construction and the Hausdorff dimension is calculated to be $d= \frac{\ln 2}{\ln 3}$~\cite{DOVGOSHEY20061}.

\begin{figure}
\centering {\ing[width=0.75\columnwidth]{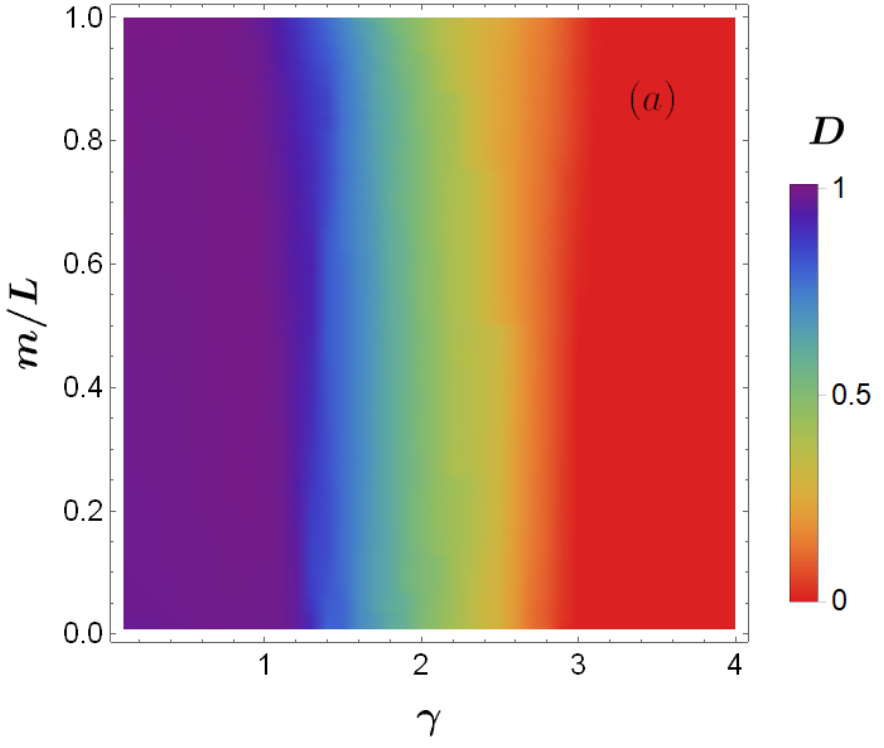}}
\centering {\ing[width=0.7\columnwidth]{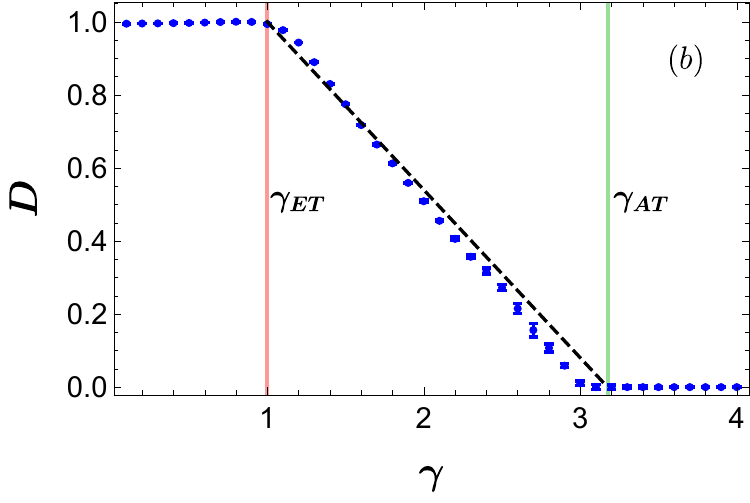}}
\caption{\imk{(a)~Energy-resolved fractal dimension $D$ vs $\gamma$ and eigenstate index $m$ (sorted in increasing order of $IPR$)
for Cantor set disorder. 
(b)~Spectral-averaged fractal dimension (blue dots) vs} $\gamma$, averaged over all eigenstates. Black dashed line indicates \imk{the analytical prediction, Eq.~\eqref{eq:Gamma_res}, for $D$ at} $1< \gamma < \gamma_{AT}$, with $ d= \frac{\ln 2}{\ln 3} \sim 0.63$. \imk{Red [green] vertical line indicates the theoretical predictions of the $\gamma_{ET}=1$  [$\gamma_{AT}$, Eq.~\eqref{eq:Andersontransition}].}}
\label{fig:cantor}
\end{figure}
In Fig.~\ref{fig:cantor}(a) we plot the second fractal dimension $D_2=D$ calculated from the numerical fitting in system size $2^p, p=7 \hdots 12$, for all the eigenvectors arranged in increasing order of IPR. In Fig.~\ref{fig:cantor}(b) we plot the same quantity, but averaged over $60$ mid spectrum states. From Fig.~\ref{fig:cantor}(a) it can be clearly seen that there is no mobility edge in the spectrum, all the eigenstates show similar fractal dimensions $D$, hence one can average over them, which is plotted in Fig.~\ref{fig:cantor}(b). The point where the system ceases to be ergodic is clearly visible at $\gamma=\gamma_{ET}=1$. Furthermore the variation of the fractal dimension of the eigenfunctions $D$ matches sufficiently well with the analytically obtained black dashed line, Eq.~\eqref{eq:Gamma_res}, in the $\gamma_{ET}<\gamma<\gamma_{AT}$ regime, thus accurately predicting the $\gamma_{AT}$ point as well. \imk{Finite-size effects, given by $1/\ln L$ terms in $D(L)$, are maximal close to $\gamma_{AT}$ and of the magnitude $\sim 0.1$.}

\paragraph*{\textbf{Generic \imk{fractal} diagonal elements:}}
Next, we consider the case of generic fractal diagonal elements. The generation of diagonal elements distributed in a generic fractal dimension was introduced recently in Ref.~\onlinecite{Kravtsov2023Cantor}, this section also serves as a demonstration of applicability of the technique. Below, we give a short summary of the method.

A random fractal spectrum of Hausdorff dimension $d$ can be generated using i.i.d. non-negative level spacings of ordered $h_{n}\leq h_{n+1}$
\be\label{eq:s-level_spacing}
s_n \equiv h_{n+1} - h_n \lra h_n = h_0 +\sum_{k=0}^{n-1} s_{k}
\ee
which \imk{are} distributed as a Pareto distribution~\cite{Arnold_1983}
\be\label{eq:P (s)_prob_level-spacing}
P (s) = \frac{d\delta_{typ}^d}{s^{d+1}}\theta (s-\delta_{typ}) \ ,
\ee
where $\delta_{typ}\sim L^{-1/d}$, is the typical level spacing of the model and we omit the subscript $n$ for brevity. 
Indeed, one can count that for the usual Cantor set with $d = \ln 2 / \ln 3$, at $n^{{\rm th}}$ step one keeps $L\cdot P (s)\sim 2^n$ levels with the spacings $s\sim 3^{-n}$, leading to the above expression. 
Due to the formal divergence of the mean level spacing 
for all $d<1$ at large $s$, for any finite $L$ one should put an upper cutoff $s_{\max}~O(1)$, given by the entire bandwidth:
\be
\delta=\mean{s}\sim \int_{\delta_{typ}}^{s_{\max}} s P (s) {\rm d}s \sim \delta_{typ}^d\sim L^{-1} \ 
\ee
and consider a typical realization where there is the only $s_n \simeq s_{\max}\simeq O(1)$, determining the bandwidth.

\begin{figure}
\centering {\ing[width=0.48\linewidth,height=0.36\columnwidth]{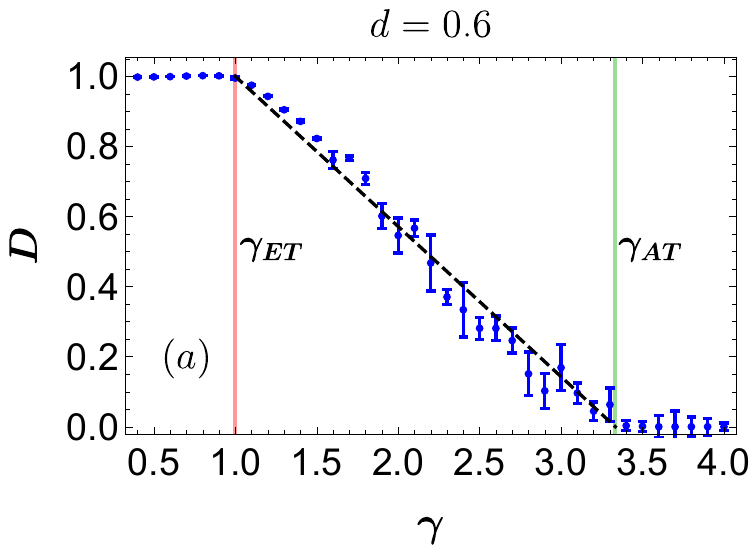}}
\centering {\ing[width=0.48\linewidth,height=0.36\columnwidth]{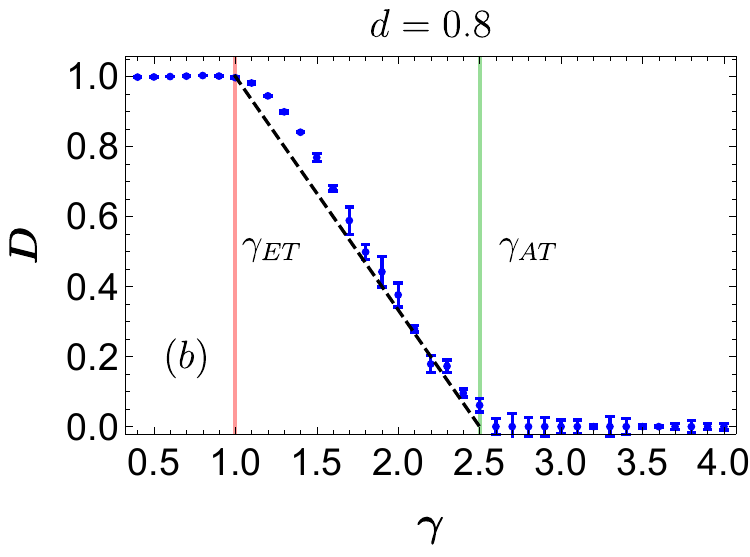}}
\caption{\imk{Fractal dimension $D$ vs $\gamma$, averaged over all eigenstates, for generic fractal diagonal disorder with Hausdorff dimension (a)~$d= 0.6$ and (b)~$d=0.8$.} \imk{The lines 
are same as in} Fig.~\ref{fig:cantor}.}
\label{fig:gen_frac}
\end{figure}
In Fig.~\ref{fig:gen_frac}(a) and~(b) we demonstrate how our theoretical predictions of $D$ match with numerical results for $d=0.6$ and $d=0.8$. We see that even for generic dimensions our analytical predictions match very well with numerics.

\paragraph*{\textbf{Multifractal disorder:}}
As a final example, we consider the more general case of multifractal
disorder. Unlike the fractal case, where the scaling behaviour of all the moments of the distribution are the same, in a multifractal they are a nontrivial function of the moment order. Thus, one needs to define the probability distribution of level spacings in an energy window appropriately scaling with system size. In this case the probability distribution of level spacings is given by~\imk{\cite{SM},}
\be\label{eq:P (s)_g (nu)_prefactor}
P\lrp{s\sim L^{-\nu}}ds = \sqrt{\frac{|g'' (\nu_0)|\ln L}{2\pi}} L^{g (\nu)-1}d\nu \ 
\ee
where $g(\nu)$ is a non-linear function of $\nu$. As an example we consider \imk{a} particular case of the log-normal distribution where, $g (\nu) = 1 - \frac{ (\nu-\nu_0)^2}{4 (\nu_0-1)}$.

Then we can compute the fractal dimension $D$ \imk{(see~\cite{SM})} as,
\begin{widetext}
\begin{gather}
D (\gamma) = \left\{
\begin{array}{cc}
1 \ , & \gamma<1 \\
2-\gamma \ , & 1<\gamma<3-\nu_0 \\
\gamma + 6\nu_0-8 -4\sqrt{ (\nu_0-1) (\gamma+2\nu_0 - 4)} \ , & 3-\nu_0<\gamma<2\nu_0\\
0 & \gamma>2\nu_0
\end{array}
\right. \label{eqn:mfrac}
\end{gather}
\end{widetext}
where the Anderson transition happens at $D = 0$, i.e., at $b = \nu_0$ and $\gamma = 2\nu_0$.
The above formula works for $1<\nu_0<2$.
Note that unlike the fractal case, here there are $4$ regimes: 
(i)~ergodic phase, $\Gamma\gg O(1)$; 
(ii)~usual fractal case, $\delta\ll \Gamma\ll O(1)$;
(ii)~new fractal case, $\delta_{typ}\ll \Gamma\ll \delta$;
(iv)~localized phase, $\Gamma\ll \delta_{typ}$.
Here the unusual fractal phase (iii) appears only when the mean level spacing $\delta$ converges and differs from the typical one, $\delta_{typ}\ll\delta\ll O(1)$.

\begin{figure}
\centering {\ing[width=0.85\linewidth,height=0.55\columnwidth]{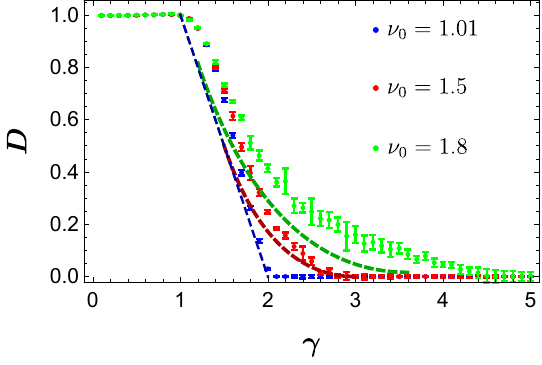}}
\caption{\imk{$D(\gamma)$ vs $\gamma$ for the log-normal disorder, Eq.~\eqref{eq:P (s)_g (nu)_prefactor}, with} $\nu_0 =1.01$ (blue), $\nu_0 = 1.5$ (red), $\nu_0 = 1.8$ (green). The dotted lines are the analytical \imk{predictions of $D (\gamma)$} from Eq.~\eqref{eqn:mfrac}.}
\label{fig:d2mfrac}
\end{figure}

The results are plotted in Fig.~\ref{fig:d2mfrac} where the predicted fractal dimensions from our saddle point approximation match well with exact numerical results. We clearly see a curvature in $D$ vs $\gamma$, a feature absent in the fractal case, which increases with increasing $\nu_0$. However it seems that 
\imk{finite-size effects} are stronger in this case than for fractal disorder, due to the logarithmic dependence of the prefactors in Eq.~\eqref{eq:P (s)_g (nu)_prefactor}\imk{, see similar effects in~\cite{LN-RP_WE,LN-RP_RRG,LN-RP_dyn}. Indeed, here finite-size effects to \imk{$D(L)$} are given by $\ln \ln L/\ln L$, that for available system sizes give $2.5$ times larger deviations.}

\paragraph*{\textbf{Level statistics:}}\imk{Until now, our focus has been exclusively on the properties of the eigenfunctions. To provide a complete analysis, we shall now study the behaviour of a signature of the phase transition in the energy levels, the consecutive level spacing ratio $r$, defined by, }
\begin{equation}
 r=\frac{{\rm min} (\delta_n,\delta_{n+1})}{{\rm max} (\delta_n,\delta_{n+1})}
 \label{eqn:r}
\end{equation}
where $\delta_n=E_n-E_{n+1}$, $E_n$ is the $n^{th}$ eigenvalue when they \imk{are} sorted in the increasing order.
In the ergodic phase, it is well known that \imk{for the Gaussian Orthogonal Ensemble (GOE) $\langle r \rangle \sim 0.53$,~\cite{oganesyan2007localization,Atas2013distribution}, where} $p (r)=\frac{27}{4}\frac{r+r^2}{ (1+r+r^2)^{5/2}}$. Deep in the localized phase we \imk{analytically derive that}
\begin{gather}
 p (r)=\frac{d}{r^{1-d}} \text{, which gives }  \langle r \rangle = \frac{d}{d+1} \ . 
 \label{eq:r_fractal}
\end{gather}
\begin{figure}
\centering {\ing[width=0.48\linewidth,height=0.36\columnwidth]{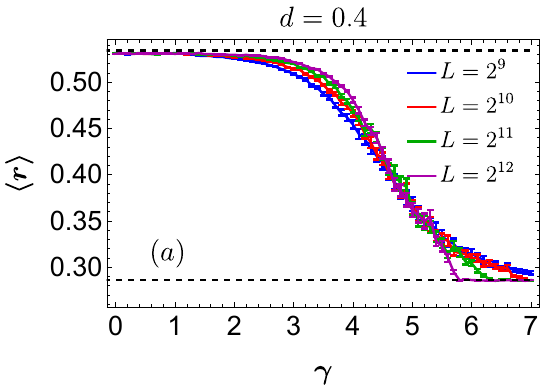}}
\centering {\ing[width=0.48\linewidth,height=0.36\columnwidth]{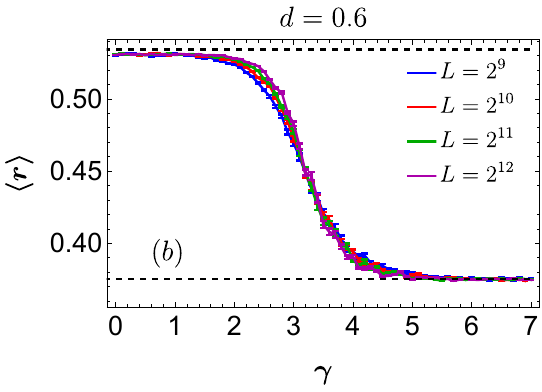}}
\centering {\ing[width=0.48\linewidth,height=0.36\columnwidth]{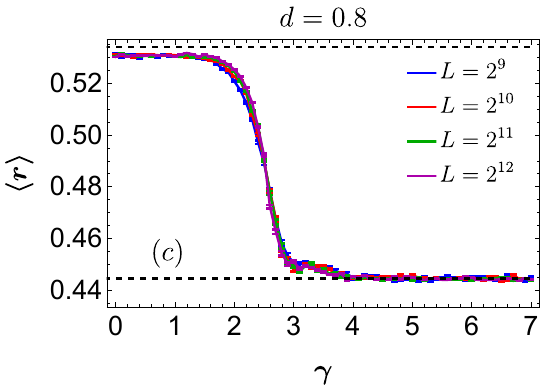}}
\centering {\ing[width=0.48\linewidth,height=0.36\columnwidth]{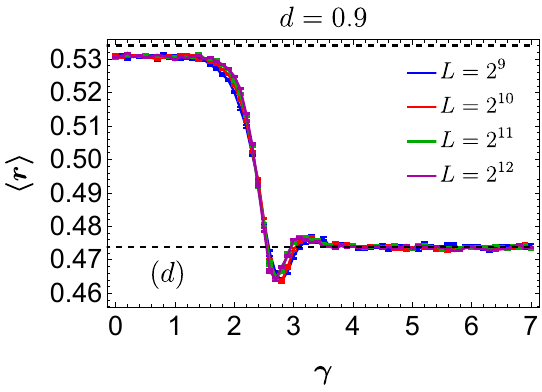}}
\caption{\imk{$\langle r \rangle$-statistics vs $\gamma$, averaged over the entire spectrum for fractal disorder with (a)~$d=0.4$, (b)~$d=0.6$, (c)~$d=0.8$, (d)~$d=0.9$, and} different system sizes $L$. The black dashed lines denote the expected values of $\langle r \rangle$ for GOE statistics and localized phase\imk{, Eq.~\eqref{eq:r_fractal}}. }
\label{fig:rstat}
\end{figure}
In Fig.~\ref{fig:rstat} we plot the variation of $\langle r \rangle$ with $\gamma$ for different $d$. As expected from our analysis for $\gamma < \gamma_{ET}=1$, it admits value close to $0.53$, while at large $\gamma >\gamma_{AT}=2/d$, it settles at $\sim \frac{d}{d+1}$, Eq.~\eqref{eq:r_fractal}. \imk{It admits intermediate values in the fractal regime, and the span in $\gamma$ where such values are observed increases with smaller $d$, consistent with our previous results.}
As the smaller $d$-values corresponds to the fatter distribution tail Eq.~\eqref{eq:P (s)_prob_level-spacing}, the finite size effects are stronger.

According to Eq.~\eqref{eq:psi_Lorenz}, $\gamma$-dependence of $\mean{r}$ goes to a kink at $\gamma=\gamma_{AT}$ in the thermodynamic limit. \imk{Interesting to note here that the fractal value $r=d/(d+1)$ covers the range from $0$ (well below Poisson value at $d=0$) to $0.5$ (rather close to GOE one at $d\to 1$).
This means that if in some other models the fractal spectrum emerges, it can be mistakenly associated with the Poisson, Wigner-Dyson or any other statistics, based solely on $r$-statistics.
Another} interesting aspect is another `kink', observed in the plots for $d\gtrsim 0.9$. While the first kink is due to breakdown of level repulsion, the second kink occurs due to the fat tail of $P(r)$ in the localized phase for $d \sim 1$. When the weight of large $r$ values for non-hybridized eigenstates deep inside the localized phase become significantly larger than what it was in the ergodic or fractal phase, it shows up as a slight increase in $\langle r \rangle$. (Also see~\cite{SM})

\paragraph*{\textbf{Discussion:}} In this work, we have demonstrated that making the distribution of the diagonal elements to be fractal in the RP model allows one to adjust the phase diagram and change the location of the Anderson localization transition $\gamma_{AT}$. We have derived an analytical expression Eq.~\eqref{eq:Gamma_res} that relates the Hausdorff dimension of the disorder to the fractal dimension of the eigenstates in the RP Hamiltonian, and have confirmed our findings through exact numerical computations. Furthermore, we have shown that one can manipulate the disorder dependence of the fractal dimension 
by utilizing a multifractal disorder. Finally, we have evaluated the implications of our modification on the eigenspectrum through level spacing ratio.

This work gives the first step in the direction of usage of the fractal disorder for the controllable tunability of the phase diagrams of various disordered models.

In particular, this work opens the way to study, whether such fractal diagonal disorder enhances fractality of wave functions in other long-range models, such as the power law banded models~\cite{MirFyod1996}, Burin-Maksimov model~\cite{Burin1989,Deng2018duality,Nosov2019correlation,deng2020anisotropy}, some Bethe-ansatz integrable ones~\cite{Richardson1963restricted,Richardson1964exact,Modak2016integrals}, on the random graphs~\cite{RRGAnnals,Tikh-Mir_RRG_review}, or even in the interacting disordered models~\cite{Abanin_RMP}.
In all these cases (especially in the latter two), the fractal disorder may open a room for non-ergodic spatially extended phase of matter, intensively discussed and highly relevant for quantum algorithms~\cite{Smelyansky_Grover} and machine learning~\cite{Smelyansky_ML}. 
The analysis of spectral statistics for $d \sim 1$ using spectral form factor can also show interesting behaviour at different timescales near $\gamma_{AT}$, which can help to identify more clearly the origin of the sudden dip in the $\langle r \rangle$ statistics and shed light on spectral distribution in the critical (fractal) regime of ergodic-localized phase transitions and structure of fractal minibands~\cite{Kravtsov2023Cantor}.

\begin{acknowledgements}
I.~M.~K. acknowledges the support
by the European Research Council under the European
Union’s Seventh Framework Program Synergy
ERC-2018-SyG HERO-810451. M.S. acknowledges support of the projects J1-2463 of the Slovenian Research Agency and EU via QuantERA grant T-NiSQ. R.G. acknowledges support from UKRI Grant
No. EP/R029075/1.
\end{acknowledgements}

\bibliography{Lib}

\begin{thebibliography}{63}%
\makeatletter
\providecommand \@ifxundefined [1]{%
 \@ifx{#1\undefined}
}%
\providecommand \@ifnum [1]{%
 \ifnum #1\expandafter \@firstoftwo
 \else \expandafter \@secondoftwo
 \fi
}%
\providecommand \@ifx [1]{%
 \ifx #1\expandafter \@firstoftwo
 \else \expandafter \@secondoftwo
 \fi
}%
\providecommand \natexlab [1]{#1}%
\providecommand \enquote  [1]{``#1''}%
\providecommand \bibnamefont  [1]{#1}%
\providecommand \bibfnamefont [1]{#1}%
\providecommand \citenamefont [1]{#1}%
\providecommand \href@noop [0]{\@secondoftwo}%
\providecommand \href [0]{\begingroup \@sanitize@url \@href}%
\providecommand \@href[1]{\@@startlink{#1}\@@href}%
\providecommand \@@href[1]{\endgroup#1\@@endlink}%
\providecommand \@sanitize@url [0]{\catcode `\\12\catcode `\$12\catcode
  `\&12\catcode `\#12\catcode `\^12\catcode `\_12\catcode `\%12\relax}%
\providecommand \@@startlink[1]{}%
\providecommand \@@endlink[0]{}%
\providecommand \url  [0]{\begingroup\@sanitize@url \@url }%
\providecommand \@url [1]{\endgroup\@href {#1}{\urlprefix }}%
\providecommand \urlprefix  [0]{URL }%
\providecommand \Eprint [0]{\href }%
\providecommand \doibase [0]{https://doi.org/}%
\providecommand \selectlanguage [0]{\@gobble}%
\providecommand \bibinfo  [0]{\@secondoftwo}%
\providecommand \bibfield  [0]{\@secondoftwo}%
\providecommand \translation [1]{[#1]}%
\providecommand \BibitemOpen [0]{}%
\providecommand \bibitemStop [0]{}%
\providecommand \bibitemNoStop [0]{.\EOS\space}%
\providecommand \EOS [0]{\spacefactor3000\relax}%
\providecommand \BibitemShut  [1]{\csname bibitem#1\endcsname}%
\let\auto@bib@innerbib\@empty
\bibitem [{\citenamefont {Evers}\ and\ \citenamefont
  {Mirlin}(2008)}]{Evers2008Anderson}%
  \BibitemOpen
  \bibfield  {author} {\bibinfo {author} {\bibfnamefont {F.}~\bibnamefont
  {Evers}}\ and\ \bibinfo {author} {\bibfnamefont {A.~D.}\ \bibnamefont
  {Mirlin}},\ }\bibfield  {title} {\bibinfo {title} {{Anderson} transitions},\
  }\href {https://doi.org/10.1103/RevModPhys.80.1355} {\bibfield  {journal}
  {\bibinfo  {journal} {Rev. Mod. Phys}\ }\textbf {\bibinfo {volume} {80}},\
  \bibinfo {pages} {1355} (\bibinfo {year} {2008})}\BibitemShut {NoStop}%
\bibitem [{\citenamefont {Basko}\ \emph {et~al.}(2006)\citenamefont {Basko},
  \citenamefont {Aleiner},\ and\ \citenamefont {Altshuler}}]{Basko06}%
  \BibitemOpen
  \bibfield  {author} {\bibinfo {author} {\bibfnamefont {D.}~\bibnamefont
  {Basko}}, \bibinfo {author} {\bibfnamefont {I.}~\bibnamefont {Aleiner}},\
  and\ \bibinfo {author} {\bibfnamefont {B.}~\bibnamefont {Altshuler}},\
  }\bibfield  {title} {\bibinfo {title} {Metal-insulator transition in a weakly
  interacting many-electron system with localized single-particle states},\
  }\href {https://doi.org/https://doi.org/10.1016/j.aop.2005.11.014} {\bibfield
   {journal} {\bibinfo  {journal} {Annals of Physics}\ }\textbf {\bibinfo
  {volume} {321}},\ \bibinfo {pages} {1126 } (\bibinfo {year}
  {2006})}\BibitemShut {NoStop}%
\bibitem [{\citenamefont {Gornyi}\ \emph {et~al.}(2005)\citenamefont {Gornyi},
  \citenamefont {Mirlin},\ and\ \citenamefont
  {Polyakov}}]{gornyi2005interacting}%
  \BibitemOpen
  \bibfield  {author} {\bibinfo {author} {\bibfnamefont {I.~V.}\ \bibnamefont
  {Gornyi}}, \bibinfo {author} {\bibfnamefont {A.~D.}\ \bibnamefont {Mirlin}},\
  and\ \bibinfo {author} {\bibfnamefont {D.~G.}\ \bibnamefont {Polyakov}},\
  }\bibfield  {title} {\bibinfo {title} {Interacting electrons in disordered
  wires: {Anderson} localization and low-$t$ transport},\ }\href
  {https://doi.org/10.1103/PhysRevLett.95.206603} {\bibfield  {journal}
  {\bibinfo  {journal} {Phys. Rev. Lett.}\ }\textbf {\bibinfo {volume} {95}},\
  \bibinfo {pages} {206603} (\bibinfo {year} {2005})}\BibitemShut {NoStop}%
\bibitem [{\citenamefont {Abanin}\ \emph {et~al.}(2019)\citenamefont {Abanin},
  \citenamefont {Altman}, \citenamefont {Bloch},\ and\ \citenamefont
  {Serbyn}}]{Abanin_RMP}%
  \BibitemOpen
  \bibfield  {author} {\bibinfo {author} {\bibfnamefont {D.~A.}\ \bibnamefont
  {Abanin}}, \bibinfo {author} {\bibfnamefont {E.}~\bibnamefont {Altman}},
  \bibinfo {author} {\bibfnamefont {I.}~\bibnamefont {Bloch}},\ and\ \bibinfo
  {author} {\bibfnamefont {M.}~\bibnamefont {Serbyn}},\ }\bibfield  {title}
  {\bibinfo {title} {Colloquium: Many-body localization, thermalization, and
  entanglement},\ }\href {https://doi.org/10.1103/RevModPhys.91.021001}
  {\bibfield  {journal} {\bibinfo  {journal} {Rev. Mod. Phys.}\ }\textbf
  {\bibinfo {volume} {91}},\ \bibinfo {pages} {021001} (\bibinfo {year}
  {2019})}\BibitemShut {NoStop}%
\bibitem [{\citenamefont {Deutsch}(1991)}]{Deutsch1991}%
  \BibitemOpen
  \bibfield  {author} {\bibinfo {author} {\bibfnamefont {J.~M.}\ \bibnamefont
  {Deutsch}},\ }\bibfield  {title} {\bibinfo {title} {Quantum statistical
  mechanics in a closed system},\ }\href
  {https://doi.org/10.1103/PhysRevA.43.2046} {\bibfield  {journal} {\bibinfo
  {journal} {Phys. Rev. A}\ }\textbf {\bibinfo {volume} {43}},\ \bibinfo
  {pages} {2046} (\bibinfo {year} {1991})}\BibitemShut {NoStop}%
\bibitem [{\citenamefont {Srednicki}(1994)}]{Srednicki1994}%
  \BibitemOpen
  \bibfield  {author} {\bibinfo {author} {\bibfnamefont {M.}~\bibnamefont
  {Srednicki}},\ }\bibfield  {title} {\bibinfo {title} {Chaos and quantum
  thermalization},\ }\href {https://doi.org/10.1103/PhysRevE.50.888} {\bibfield
   {journal} {\bibinfo  {journal} {Phys. Rev. E}\ }\textbf {\bibinfo {volume}
  {50}},\ \bibinfo {pages} {888} (\bibinfo {year} {1994})}\BibitemShut
  {NoStop}%
\bibitem [{\citenamefont {Luitz}\ \emph {et~al.}(2015)\citenamefont {Luitz},
  \citenamefont {Laflorencie},\ and\ \citenamefont {Alet}}]{Luitz15}%
  \BibitemOpen
  \bibfield  {author} {\bibinfo {author} {\bibfnamefont {D.~J.}\ \bibnamefont
  {Luitz}}, \bibinfo {author} {\bibfnamefont {N.}~\bibnamefont {Laflorencie}},\
  and\ \bibinfo {author} {\bibfnamefont {F.}~\bibnamefont {Alet}},\ }\bibfield
  {title} {\bibinfo {title} {Many-body localization edge in the random-field
  {H}eisenberg chain},\ }\href {https://doi.org/10.1103/PhysRevB.91.081103}
  {\bibfield  {journal} {\bibinfo  {journal} {Phys. Rev. B}\ }\textbf {\bibinfo
  {volume} {91}},\ \bibinfo {pages} {081103} (\bibinfo {year}
  {2015})}\BibitemShut {NoStop}%
\bibitem [{\citenamefont {Mac\'e}\ \emph {et~al.}(2019)\citenamefont {Mac\'e},
  \citenamefont {Alet},\ and\ \citenamefont
  {Laflorencie}}]{Mace_Laflorencie2019_XXZ}%
  \BibitemOpen
  \bibfield  {author} {\bibinfo {author} {\bibfnamefont {N.}~\bibnamefont
  {Mac\'e}}, \bibinfo {author} {\bibfnamefont {F.}~\bibnamefont {Alet}},\ and\
  \bibinfo {author} {\bibfnamefont {N.}~\bibnamefont {Laflorencie}},\
  }\bibfield  {title} {\bibinfo {title} {Multifractal scalings across the
  many-body localization transition},\ }\href
  {https://doi.org/10.1103/PhysRevLett.123.180601} {\bibfield  {journal}
  {\bibinfo  {journal} {Phys. Rev. Lett.}\ }\textbf {\bibinfo {volume} {123}},\
  \bibinfo {pages} {180601} (\bibinfo {year} {2019})}\BibitemShut {NoStop}%
\bibitem [{\citenamefont {De~Tomasi}\ \emph {et~al.}(2021)\citenamefont
  {De~Tomasi}, \citenamefont {Khaymovich}, \citenamefont {Pollmann},\ and\
  \citenamefont {Warzel}}]{QIsing_2021}%
  \BibitemOpen
  \bibfield  {author} {\bibinfo {author} {\bibfnamefont {G.}~\bibnamefont
  {De~Tomasi}}, \bibinfo {author} {\bibfnamefont {I.~M.}\ \bibnamefont
  {Khaymovich}}, \bibinfo {author} {\bibfnamefont {F.}~\bibnamefont
  {Pollmann}},\ and\ \bibinfo {author} {\bibfnamefont {S.}~\bibnamefont
  {Warzel}},\ }\bibfield  {title} {\bibinfo {title} {Rare thermal bubbles at
  the many-body localization transition from the {Fock} space point of view},\
  }\href {https://doi.org/10.1103/PhysRevB.104.024202} {\bibfield  {journal}
  {\bibinfo  {journal} {Phys. Rev. B}\ }\textbf {\bibinfo {volume} {104}},\
  \bibinfo {pages} {024202} (\bibinfo {year} {2021})}\BibitemShut {NoStop}%
\bibitem [{\citenamefont {{Bar Lev}}\ \emph {et~al.}(2015)\citenamefont {{Bar
  Lev}}, \citenamefont {Cohen},\ and\ \citenamefont
  {Reichman}}]{BarLev2015absence}%
  \BibitemOpen
  \bibfield  {author} {\bibinfo {author} {\bibfnamefont {Y.}~\bibnamefont {{Bar
  Lev}}}, \bibinfo {author} {\bibfnamefont {G.}~\bibnamefont {Cohen}},\ and\
  \bibinfo {author} {\bibfnamefont {D.~R.}\ \bibnamefont {Reichman}},\
  }\bibfield  {title} {\bibinfo {title} {Absence of diffusion in an interacting
  system of spinless fermions on a one-dimensional disordered lattice},\ }\href
  {https://doi.org/10.1103/PhysRevLett.114.100601} {\bibfield  {journal}
  {\bibinfo  {journal} {Phys. Rev. Lett.}\ }\textbf {\bibinfo {volume} {114}},\
  \bibinfo {pages} {100601} (\bibinfo {year} {2015})}\BibitemShut {NoStop}%
\bibitem [{\citenamefont {Agarwal}\ \emph {et~al.}(2015)\citenamefont
  {Agarwal}, \citenamefont {Gopalakrishnan}, \citenamefont {Knap},
  \citenamefont {M\"uller},\ and\ \citenamefont
  {Demler}}]{Griffiths2015anomalous}%
  \BibitemOpen
  \bibfield  {author} {\bibinfo {author} {\bibfnamefont {K.}~\bibnamefont
  {Agarwal}}, \bibinfo {author} {\bibfnamefont {S.}~\bibnamefont
  {Gopalakrishnan}}, \bibinfo {author} {\bibfnamefont {M.}~\bibnamefont
  {Knap}}, \bibinfo {author} {\bibfnamefont {M.}~\bibnamefont {M\"uller}},\
  and\ \bibinfo {author} {\bibfnamefont {E.}~\bibnamefont {Demler}},\
  }\bibfield  {title} {\bibinfo {title} {Anomalous diffusion and {Griffiths}
  effects near the many-body localization transition},\ }\href
  {https://doi.org/10.1103/PhysRevLett.114.160401} {\bibfield  {journal}
  {\bibinfo  {journal} {Phys. Rev. Lett.}\ }\textbf {\bibinfo {volume} {114}},\
  \bibinfo {pages} {160401} (\bibinfo {year} {2015})}\BibitemShut {NoStop}%
\bibitem [{\citenamefont {Luitz}\ \emph {et~al.}(2016)\citenamefont {Luitz},
  \citenamefont {Laflorencie},\ and\ \citenamefont {Alet}}]{Luitz2016extended}%
  \BibitemOpen
  \bibfield  {author} {\bibinfo {author} {\bibfnamefont {D.~J.}\ \bibnamefont
  {Luitz}}, \bibinfo {author} {\bibfnamefont {N.}~\bibnamefont {Laflorencie}},\
  and\ \bibinfo {author} {\bibfnamefont {F.}~\bibnamefont {Alet}},\ }\bibfield
  {title} {\bibinfo {title} {Extended slow dynamical regime close to the
  many-body localization transition},\ }\href
  {https://doi.org/10.1103/PhysRevB.93.060201} {\bibfield  {journal} {\bibinfo
  {journal} {Phys. Rev. B}\ }\textbf {\bibinfo {volume} {93}},\ \bibinfo
  {pages} {060201} (\bibinfo {year} {2016})}\BibitemShut {NoStop}%
\bibitem [{\citenamefont {Luitz}\ and\ \citenamefont {{Bar
  Lev}}(2016)}]{Luitz2016anomalous}%
  \BibitemOpen
  \bibfield  {author} {\bibinfo {author} {\bibfnamefont {D.~J.}\ \bibnamefont
  {Luitz}}\ and\ \bibinfo {author} {\bibfnamefont {Y.}~\bibnamefont {{Bar
  Lev}}},\ }\bibfield  {title} {\bibinfo {title} {Anomalous thermalization in
  ergodic systems},\ }\href {https://doi.org/10.1103/PhysRevLett.117.170404}
  {\bibfield  {journal} {\bibinfo  {journal} {Phys. Rev. Lett.}\ }\textbf
  {\bibinfo {volume} {117}},\ \bibinfo {pages} {170404} (\bibinfo {year}
  {2016})}\BibitemShut {NoStop}%
\bibitem [{\citenamefont {Khait}\ \emph {et~al.}(2016)\citenamefont {Khait},
  \citenamefont {Gazit}, \citenamefont {Yao},\ and\ \citenamefont
  {Auerbach}}]{Khait2016spin}%
  \BibitemOpen
  \bibfield  {author} {\bibinfo {author} {\bibfnamefont {I.}~\bibnamefont
  {Khait}}, \bibinfo {author} {\bibfnamefont {S.}~\bibnamefont {Gazit}},
  \bibinfo {author} {\bibfnamefont {N.~Y.}\ \bibnamefont {Yao}},\ and\ \bibinfo
  {author} {\bibfnamefont {A.}~\bibnamefont {Auerbach}},\ }\bibfield  {title}
  {\bibinfo {title} {Spin transport of weakly disordered {Heisenberg} chain at
  infinite temperature},\ }\href {https://doi.org/10.1103/PhysRevB.93.224205}
  {\bibfield  {journal} {\bibinfo  {journal} {Phys. Rev. B}\ }\textbf {\bibinfo
  {volume} {93}},\ \bibinfo {pages} {224205} (\bibinfo {year}
  {2016})}\BibitemShut {NoStop}%
\bibitem [{\citenamefont {\ifmmode \check{Z}\else
  \v{Z}\fi{}nidari\ifmmode~\check{c}\else \v{c}\fi{}}\ \emph
  {et~al.}(2016)\citenamefont {\ifmmode \check{Z}\else
  \v{Z}\fi{}nidari\ifmmode~\check{c}\else \v{c}\fi{}}, \citenamefont
  {Scardicchio},\ and\ \citenamefont {Varma}}]{Znidaric2016Diffusive}%
  \BibitemOpen
  \bibfield  {author} {\bibinfo {author} {\bibfnamefont {M.}~\bibnamefont
  {\ifmmode \check{Z}\else \v{Z}\fi{}nidari\ifmmode~\check{c}\else
  \v{c}\fi{}}}, \bibinfo {author} {\bibfnamefont {A.}~\bibnamefont
  {Scardicchio}},\ and\ \bibinfo {author} {\bibfnamefont {V.~K.}\ \bibnamefont
  {Varma}},\ }\bibfield  {title} {\bibinfo {title} {Diffusive and subdiffusive
  spin transport in the ergodic phase of a many-body localizable system},\
  }\href {https://doi.org/10.1103/PhysRevLett.117.040601} {\bibfield  {journal}
  {\bibinfo  {journal} {Phys. Rev. Lett.}\ }\textbf {\bibinfo {volume} {117}},\
  \bibinfo {pages} {040601} (\bibinfo {year} {2016})}\BibitemShut {NoStop}%
\bibitem [{\citenamefont {{Bar Lev}}\ \emph {et~al.}(2017)\citenamefont {{Bar
  Lev}}, \citenamefont {Kennes}, \citenamefont {Klöckner}, \citenamefont
  {Reichman},\ and\ \citenamefont {Karrasch}}]{BarLev2017transport}%
  \BibitemOpen
  \bibfield  {author} {\bibinfo {author} {\bibfnamefont {Y.}~\bibnamefont {{Bar
  Lev}}}, \bibinfo {author} {\bibfnamefont {D.~M.}\ \bibnamefont {Kennes}},
  \bibinfo {author} {\bibfnamefont {C.}~\bibnamefont {Klöckner}}, \bibinfo
  {author} {\bibfnamefont {D.~R.}\ \bibnamefont {Reichman}},\ and\ \bibinfo
  {author} {\bibfnamefont {C.}~\bibnamefont {Karrasch}},\ }\bibfield  {title}
  {\bibinfo {title} {Transport in quasiperiodic interacting systems: From
  superdiffusion to subdiffusion},\ }\href
  {https://doi.org/10.1209/0295-5075/119/37003} {\bibfield  {journal} {\bibinfo
   {journal} {{EPL} (Europhysics Letters)}\ }\textbf {\bibinfo {volume}
  {119}},\ \bibinfo {pages} {37003} (\bibinfo {year} {2017})}\BibitemShut
  {NoStop}%
\bibitem [{\citenamefont {Bera}\ \emph {et~al.}(2017)\citenamefont {Bera},
  \citenamefont {{De Tomasi}}, \citenamefont {Weiner},\ and\ \citenamefont
  {Evers}}]{Bera2017density}%
  \BibitemOpen
  \bibfield  {author} {\bibinfo {author} {\bibfnamefont {S.}~\bibnamefont
  {Bera}}, \bibinfo {author} {\bibfnamefont {G.}~\bibnamefont {{De Tomasi}}},
  \bibinfo {author} {\bibfnamefont {F.}~\bibnamefont {Weiner}},\ and\ \bibinfo
  {author} {\bibfnamefont {F.}~\bibnamefont {Evers}},\ }\bibfield  {title}
  {\bibinfo {title} {Density propagator for many-body localization: Finite-size
  effects, transient subdiffusion, and exponential decay},\ }\href
  {https://doi.org/10.1103/PhysRevLett.118.196801} {\bibfield  {journal}
  {\bibinfo  {journal} {Phys. Rev. Lett.}\ }\textbf {\bibinfo {volume} {118}},\
  \bibinfo {pages} {196801} (\bibinfo {year} {2017})}\BibitemShut {NoStop}%
\bibitem [{\citenamefont {Agarwal}\ \emph {et~al.}(2017)\citenamefont
  {Agarwal}, \citenamefont {Altman}, \citenamefont {Demler}, \citenamefont
  {Gopalakrishnan}, \citenamefont {Huse},\ and\ \citenamefont
  {Knap}}]{agarwal2017rare}%
  \BibitemOpen
  \bibfield  {author} {\bibinfo {author} {\bibfnamefont {K.}~\bibnamefont
  {Agarwal}}, \bibinfo {author} {\bibfnamefont {E.}~\bibnamefont {Altman}},
  \bibinfo {author} {\bibfnamefont {E.}~\bibnamefont {Demler}}, \bibinfo
  {author} {\bibfnamefont {S.}~\bibnamefont {Gopalakrishnan}}, \bibinfo
  {author} {\bibfnamefont {D.~A.}\ \bibnamefont {Huse}},\ and\ \bibinfo
  {author} {\bibfnamefont {M.}~\bibnamefont {Knap}},\ }\bibfield  {title}
  {\bibinfo {title} {Rare-region effects and dynamics near the many-body
  localization transition},\ }\href {https://doi.org/10.1002/andp.201600326}
  {\bibfield  {journal} {\bibinfo  {journal} {Annalen der Physik}\ }\textbf
  {\bibinfo {volume} {529}},\ \bibinfo {pages} {1600326} (\bibinfo {year}
  {2017})}\BibitemShut {NoStop}%
\bibitem [{\citenamefont {Luitz}\ and\ \citenamefont {{Bar
  Lev}}(2017)}]{luitz2017ergodic}%
  \BibitemOpen
  \bibfield  {author} {\bibinfo {author} {\bibfnamefont {D.~J.}\ \bibnamefont
  {Luitz}}\ and\ \bibinfo {author} {\bibfnamefont {Y.}~\bibnamefont {{Bar
  Lev}}},\ }\bibfield  {title} {\bibinfo {title} {The ergodic side of the
  many-body localization transition},\ }\href
  {https://doi.org/10.1002/andp.201600350} {\bibfield  {journal} {\bibinfo
  {journal} {Annalen der Physik}\ }\textbf {\bibinfo {volume} {529}},\ \bibinfo
  {pages} {1600350} (\bibinfo {year} {2017})}\BibitemShut {NoStop}%
\bibitem [{\citenamefont {Lezama}\ \emph {et~al.}(2019)\citenamefont {Lezama},
  \citenamefont {Bera},\ and\ \citenamefont {Bardarson}}]{Lezama2019apparent}%
  \BibitemOpen
  \bibfield  {author} {\bibinfo {author} {\bibfnamefont {T.~L.~M.}\
  \bibnamefont {Lezama}}, \bibinfo {author} {\bibfnamefont {S.}~\bibnamefont
  {Bera}},\ and\ \bibinfo {author} {\bibfnamefont {J.~H.}\ \bibnamefont
  {Bardarson}},\ }\bibfield  {title} {\bibinfo {title} {Apparent slow dynamics
  in the ergodic phase of a driven many-body localized system without extensive
  conserved quantities},\ }\href {https://doi.org/10.1103/PhysRevB.99.161106}
  {\bibfield  {journal} {\bibinfo  {journal} {Phys. Rev. B}\ }\textbf {\bibinfo
  {volume} {99}},\ \bibinfo {pages} {161106} (\bibinfo {year}
  {2019})}\BibitemShut {NoStop}%
\bibitem [{\citenamefont {Roy}\ \emph {et~al.}(2018{\natexlab{a}})\citenamefont
  {Roy}, \citenamefont {{Bar Lev}},\ and\ \citenamefont
  {Luitz}}]{roy2018anomalous}%
  \BibitemOpen
  \bibfield  {author} {\bibinfo {author} {\bibfnamefont {S.}~\bibnamefont
  {Roy}}, \bibinfo {author} {\bibfnamefont {Y.}~\bibnamefont {{Bar Lev}}},\
  and\ \bibinfo {author} {\bibfnamefont {D.~J.}\ \bibnamefont {Luitz}},\
  }\bibfield  {title} {\bibinfo {title} {Anomalous thermalization and transport
  in disordered interacting {Floquet} systems},\ }\href
  {https://doi.org/10.1103/PhysRevB.98.060201} {\bibfield  {journal} {\bibinfo
  {journal} {Physical Review B}\ }\textbf {\bibinfo {volume} {98}},\ \bibinfo
  {pages} {060201} (\bibinfo {year} {2018}{\natexlab{a}})}\BibitemShut
  {NoStop}%
\bibitem [{\citenamefont {Roy}\ \emph {et~al.}(2018{\natexlab{b}})\citenamefont
  {Roy}, \citenamefont {Khaymovich}, \citenamefont {Das},\ and\ \citenamefont
  {Moessner}}]{Floquet_MF}%
  \BibitemOpen
  \bibfield  {author} {\bibinfo {author} {\bibfnamefont {S.}~\bibnamefont
  {Roy}}, \bibinfo {author} {\bibfnamefont {I.~M.}\ \bibnamefont {Khaymovich}},
  \bibinfo {author} {\bibfnamefont {A.}~\bibnamefont {Das}},\ and\ \bibinfo
  {author} {\bibfnamefont {R.}~\bibnamefont {Moessner}},\ }\bibfield  {title}
  {\bibinfo {title} {{Multifractality without fine-tuning in a {Floquet}
  quasiperiodic chain}},\ }\href {https://doi.org/10.21468/SciPostPhys.4.5.025}
  {\bibfield  {journal} {\bibinfo  {journal} {SciPost Phys.}\ }\textbf
  {\bibinfo {volume} {4}},\ \bibinfo {pages} {25} (\bibinfo {year}
  {2018}{\natexlab{b}})}\BibitemShut {NoStop}%
\bibitem [{\citenamefont {Sarkar}\ \emph {et~al.}(2021)\citenamefont {Sarkar},
  \citenamefont {Ghosh}, \citenamefont {Sen},\ and\ \citenamefont
  {Sengupta}}]{Sarkarnonintmfrac}%
  \BibitemOpen
  \bibfield  {author} {\bibinfo {author} {\bibfnamefont {M.}~\bibnamefont
  {Sarkar}}, \bibinfo {author} {\bibfnamefont {R.}~\bibnamefont {Ghosh}},
  \bibinfo {author} {\bibfnamefont {A.}~\bibnamefont {Sen}},\ and\ \bibinfo
  {author} {\bibfnamefont {K.}~\bibnamefont {Sengupta}},\ }\bibfield  {title}
  {\bibinfo {title} {Mobility edge and multifractality in a periodically driven
  {Aubry}-{Andr\'e} model},\ }\href
  {https://doi.org/10.1103/PhysRevB.103.184309} {\bibfield  {journal} {\bibinfo
   {journal} {Phys. Rev. B}\ }\textbf {\bibinfo {volume} {103}},\ \bibinfo
  {pages} {184309} (\bibinfo {year} {2021})}\BibitemShut {NoStop}%
\bibitem [{\citenamefont {Sarkar}\ \emph {et~al.}(2022)\citenamefont {Sarkar},
  \citenamefont {Ghosh}, \citenamefont {Sen},\ and\ \citenamefont
  {Sengupta}}]{Sarkarintmfrac}%
  \BibitemOpen
  \bibfield  {author} {\bibinfo {author} {\bibfnamefont {M.}~\bibnamefont
  {Sarkar}}, \bibinfo {author} {\bibfnamefont {R.}~\bibnamefont {Ghosh}},
  \bibinfo {author} {\bibfnamefont {A.}~\bibnamefont {Sen}},\ and\ \bibinfo
  {author} {\bibfnamefont {K.}~\bibnamefont {Sengupta}},\ }\bibfield  {title}
  {\bibinfo {title} {Signatures of multifractality in a periodically driven
  interacting {Aubry}-{Andr\'e} model},\ }\href
  {https://doi.org/10.1103/PhysRevB.105.024301} {\bibfield  {journal} {\bibinfo
   {journal} {Phys. Rev. B}\ }\textbf {\bibinfo {volume} {105}},\ \bibinfo
  {pages} {024301} (\bibinfo {year} {2022})}\BibitemShut {NoStop}%
\bibitem [{\citenamefont {Aditya}\ \emph {et~al.}(2023)\citenamefont {Aditya},
  \citenamefont {Sengupta},\ and\ \citenamefont
  {Sen}}]{SenPhysRevB.107.035402}%
  \BibitemOpen
  \bibfield  {author} {\bibinfo {author} {\bibfnamefont {S.}~\bibnamefont
  {Aditya}}, \bibinfo {author} {\bibfnamefont {K.}~\bibnamefont {Sengupta}},\
  and\ \bibinfo {author} {\bibfnamefont {D.}~\bibnamefont {Sen}},\ }\bibfield
  {title} {\bibinfo {title} {Periodically driven model with quasiperiodic
  potential and staggered hopping amplitudes: Engineering of mobility gaps and
  multifractal states},\ }\href {https://doi.org/10.1103/PhysRevB.107.035402}
  {\bibfield  {journal} {\bibinfo  {journal} {Phys. Rev. B}\ }\textbf {\bibinfo
  {volume} {107}},\ \bibinfo {pages} {035402} (\bibinfo {year}
  {2023})}\BibitemShut {NoStop}%
\bibitem [{\citenamefont {Deng}\ \emph {et~al.}(2019)\citenamefont {Deng},
  \citenamefont {Ray}, \citenamefont {Sinha}, \citenamefont {Shlyapnikov},\
  and\ \citenamefont {Santos}}]{SayakPhysRevLett.123.025301}%
  \BibitemOpen
  \bibfield  {author} {\bibinfo {author} {\bibfnamefont {X.}~\bibnamefont
  {Deng}}, \bibinfo {author} {\bibfnamefont {S.}~\bibnamefont {Ray}}, \bibinfo
  {author} {\bibfnamefont {S.}~\bibnamefont {Sinha}}, \bibinfo {author}
  {\bibfnamefont {G.~V.}\ \bibnamefont {Shlyapnikov}},\ and\ \bibinfo {author}
  {\bibfnamefont {L.}~\bibnamefont {Santos}},\ }\bibfield  {title} {\bibinfo
  {title} {One-dimensional quasicrystals with power-law hopping},\ }\href
  {https://doi.org/10.1103/PhysRevLett.123.025301} {\bibfield  {journal}
  {\bibinfo  {journal} {Phys. Rev. Lett.}\ }\textbf {\bibinfo {volume} {123}},\
  \bibinfo {pages} {025301} (\bibinfo {year} {2019})}\BibitemShut {NoStop}%
\bibitem [{\citenamefont {Serbyn}\ \emph {et~al.}(2017)\citenamefont {Serbyn},
  \citenamefont {Papi\ifmmode~\acute{c}\else \'{c}\fi{}},\ and\ \citenamefont
  {Abanin}}]{AbaninPhysRevB.96.104201}%
  \BibitemOpen
  \bibfield  {author} {\bibinfo {author} {\bibfnamefont {M.}~\bibnamefont
  {Serbyn}}, \bibinfo {author} {\bibfnamefont {Z.}~\bibnamefont
  {Papi\ifmmode~\acute{c}\else \'{c}\fi{}}},\ and\ \bibinfo {author}
  {\bibfnamefont {D.~A.}\ \bibnamefont {Abanin}},\ }\bibfield  {title}
  {\bibinfo {title} {Thouless energy and multifractality across the many-body
  localization transition},\ }\href
  {https://doi.org/10.1103/PhysRevB.96.104201} {\bibfield  {journal} {\bibinfo
  {journal} {Phys. Rev. B}\ }\textbf {\bibinfo {volume} {96}},\ \bibinfo
  {pages} {104201} (\bibinfo {year} {2017})}\BibitemShut {NoStop}%
\bibitem [{\citenamefont {{Rosenzweig}}\ and\ \citenamefont
  {{Porter}}(1960)}]{RP}%
  \BibitemOpen
  \bibfield  {author} {\bibinfo {author} {\bibfnamefont {N.}~\bibnamefont
  {{Rosenzweig}}}\ and\ \bibinfo {author} {\bibfnamefont {C.~E.}\ \bibnamefont
  {{Porter}}},\ }\bibfield  {title} {\bibinfo {title} {"{R}epulsion of energy
  levels" in complex atomic spectra},\ }\href
  {https://doi.org/10.1103/PhysRev.120.1698} {\bibfield  {journal} {\bibinfo
  {journal} {Phys. Rev. B}\ }\textbf {\bibinfo {volume} {120}},\ \bibinfo
  {pages} {1698} (\bibinfo {year} {1960})}\BibitemShut {NoStop}%
\bibitem [{\citenamefont {Kravtsov}\ \emph {et~al.}(2015)\citenamefont
  {Kravtsov}, \citenamefont {Khaymovich}, \citenamefont {Cuevas},\ and\
  \citenamefont {Amini}}]{Kravtsov_NJP2015}%
  \BibitemOpen
  \bibfield  {author} {\bibinfo {author} {\bibfnamefont {V.~E.}\ \bibnamefont
  {Kravtsov}}, \bibinfo {author} {\bibfnamefont {I.~M.}\ \bibnamefont
  {Khaymovich}}, \bibinfo {author} {\bibfnamefont {E.}~\bibnamefont {Cuevas}},\
  and\ \bibinfo {author} {\bibfnamefont {M.}~\bibnamefont {Amini}},\ }\bibfield
   {title} {\bibinfo {title} {A random matrix model with localization and
  ergodic transitions},\ }\href
  {https://doi.org/10.1088/1367-2630/17/12/122002} {\bibfield  {journal}
  {\bibinfo  {journal} {New J. Phys.}\ }\textbf {\bibinfo {volume} {17}},\
  \bibinfo {pages} {122002} (\bibinfo {year} {2015})}\BibitemShut {NoStop}%
\bibitem [{\citenamefont {Facoetti}\ \emph {et~al.}(2016)\citenamefont
  {Facoetti}, \citenamefont {Vivo},\ and\ \citenamefont {Biroli}}]{Biroli_RP}%
  \BibitemOpen
  \bibfield  {author} {\bibinfo {author} {\bibfnamefont {D.}~\bibnamefont
  {Facoetti}}, \bibinfo {author} {\bibfnamefont {P.}~\bibnamefont {Vivo}},\
  and\ \bibinfo {author} {\bibfnamefont {G.}~\bibnamefont {Biroli}},\
  }\bibfield  {title} {\bibinfo {title} {From non-ergodic eigenvectors to local
  resolvent statistics and back: A random matrix perspective},\ }\href
  {https://doi.org/10.1209/0295-5075/115/47003} {\bibfield  {journal} {\bibinfo
   {journal} {Europhys. Lett.}\ }\textbf {\bibinfo {volume} {115}},\ \bibinfo
  {pages} {47003} (\bibinfo {year} {2016})}\BibitemShut {NoStop}%
\bibitem [{\citenamefont {Truong}\ and\ \citenamefont
  {Ossipov}(2016)}]{Ossipov_EPL2016_H+V}%
  \BibitemOpen
  \bibfield  {author} {\bibinfo {author} {\bibfnamefont {K.}~\bibnamefont
  {Truong}}\ and\ \bibinfo {author} {\bibfnamefont {A.}~\bibnamefont
  {Ossipov}},\ }\bibfield  {title} {\bibinfo {title} {Eigenvectors under a
  generic perturbation: Non-perturbative results from the random matrix
  approach},\ }\href {https://doi.org/10.1209/0295-5075/116/37002} {\bibfield
  {journal} {\bibinfo  {journal} {Europhys. Lett.}\ }\textbf {\bibinfo {volume}
  {116}},\ \bibinfo {pages} {37002} (\bibinfo {year} {2016})}\BibitemShut
  {NoStop}%
\bibitem [{\citenamefont {Monthus}(2017)}]{Monthus}%
  \BibitemOpen
  \bibfield  {author} {\bibinfo {author} {\bibfnamefont {C.}~\bibnamefont
  {Monthus}},\ }\bibfield  {title} {\bibinfo {title} {Statistical properties of
  the {Green} function in finite size for {Anderson} localization models with
  multifractal eigenvectors},\ }\href
  {https://doi.org/10.1088/1751-8121/aa5ad2} {\bibfield  {journal} {\bibinfo
  {journal} {J. Phys. A: Math. Theor.}\ }\textbf {\bibinfo {volume} {50}},\
  \bibinfo {pages} {295101} (\bibinfo {year} {2017})}\BibitemShut {NoStop}%
\bibitem [{\citenamefont {Bogomolny}\ and\ \citenamefont
  {Sieber}(2018)}]{BogomolnyRP2018}%
  \BibitemOpen
  \bibfield  {author} {\bibinfo {author} {\bibfnamefont {E.}~\bibnamefont
  {Bogomolny}}\ and\ \bibinfo {author} {\bibfnamefont {M.}~\bibnamefont
  {Sieber}},\ }\bibfield  {title} {\bibinfo {title} {Eigenfunction distribution
  for the {Rosenzweig}-{Porter} model},\ }\href
  {https://doi.org/10.1103/PhysRevE.98.032139} {\bibfield  {journal} {\bibinfo
  {journal} {Phys. Rev. E}\ }\textbf {\bibinfo {volume} {98}},\ \bibinfo
  {pages} {032139} (\bibinfo {year} {2018})}\BibitemShut {NoStop}%
\bibitem [{\citenamefont {Venturelli}\ \emph {et~al.}(2023)\citenamefont
  {Venturelli}, \citenamefont {Cugliandolo}, \citenamefont {Schehr},\ and\
  \citenamefont {Tarzia}}]{Venturelli2023replica}%
  \BibitemOpen
  \bibfield  {author} {\bibinfo {author} {\bibfnamefont {D.}~\bibnamefont
  {Venturelli}}, \bibinfo {author} {\bibfnamefont {L.~F.}\ \bibnamefont
  {Cugliandolo}}, \bibinfo {author} {\bibfnamefont {G.}~\bibnamefont
  {Schehr}},\ and\ \bibinfo {author} {\bibfnamefont {M.}~\bibnamefont
  {Tarzia}},\ }\bibfield  {title} {\bibinfo {title} {{Replica approach to the
  generalized {Rosenzweig-Porter} model}},\ }\href
  {https://doi.org/10.21468/SciPostPhys.14.5.110} {\bibfield  {journal}
  {\bibinfo  {journal} {SciPost Phys.}\ }\textbf {\bibinfo {volume} {14}},\
  \bibinfo {pages} {110} (\bibinfo {year} {2023})}\BibitemShut {NoStop}%
\bibitem [{\citenamefont {von Soosten}\ and\ \citenamefont
  {Warzel}(2018)}]{vonSoosten2017non}%
  \BibitemOpen
  \bibfield  {author} {\bibinfo {author} {\bibfnamefont {P.}~\bibnamefont {von
  Soosten}}\ and\ \bibinfo {author} {\bibfnamefont {S.}~\bibnamefont
  {Warzel}},\ }\bibfield  {title} {\bibinfo {title} {Non-ergodic delocalization
  in the {Rosenzweig}--{Porter} model},\ }\href
  {https://doi.org/10.1007/s11005-018-1131-7} {\bibfield  {journal} {\bibinfo
  {journal} {Letters in Mathematical Physics}\ ,\ \bibinfo {pages} {1}}
  (\bibinfo {year} {2018})}\BibitemShut {NoStop}%
\bibitem [{\citenamefont {Kravtsov}\ \emph {et~al.}(2020)\citenamefont
  {Kravtsov}, \citenamefont {Khaymovich}, \citenamefont {Altshuler},\ and\
  \citenamefont {Ioffe}}]{LN-RP_RRG}%
  \BibitemOpen
  \bibfield  {author} {\bibinfo {author} {\bibfnamefont {V.~E.}\ \bibnamefont
  {Kravtsov}}, \bibinfo {author} {\bibfnamefont {I.~M.}\ \bibnamefont
  {Khaymovich}}, \bibinfo {author} {\bibfnamefont {B.~L.}\ \bibnamefont
  {Altshuler}},\ and\ \bibinfo {author} {\bibfnamefont {L.~B.}\ \bibnamefont
  {Ioffe}},\ }\bibfield  {title} {\bibinfo {title} {Localization transition on
  the random regular graph as an unstable tricritical point in a log-normal
  {Rosenzweig}-{Porter} random matrix ensemble},\ }\Eprint
  {https://arxiv.org/abs/2002.02979} {arXiv:2002.02979}  (\bibinfo {year}
  {2020})\BibitemShut {NoStop}%
\bibitem [{\citenamefont {Khaymovich}\ \emph {et~al.}(2020)\citenamefont
  {Khaymovich}, \citenamefont {Kravtsov}, \citenamefont {Altshuler},\ and\
  \citenamefont {Ioffe}}]{LN-RP_WE}%
  \BibitemOpen
  \bibfield  {author} {\bibinfo {author} {\bibfnamefont {I.~M.}\ \bibnamefont
  {Khaymovich}}, \bibinfo {author} {\bibfnamefont {V.~E.}\ \bibnamefont
  {Kravtsov}}, \bibinfo {author} {\bibfnamefont {B.~L.}\ \bibnamefont
  {Altshuler}},\ and\ \bibinfo {author} {\bibfnamefont {L.~B.}\ \bibnamefont
  {Ioffe}},\ }\bibfield  {title} {\bibinfo {title} {Fragile ergodic phases in
  logarithmically-normal {Rosenzweig}-{{Porter}} model},\ }\href
  {https://doi.org/10.1103/PhysRevResearch.2.043346} {\bibfield  {journal}
  {\bibinfo  {journal} {Phys. Rev. Research}\ }\textbf {\bibinfo {volume}
  {2}},\ \bibinfo {pages} {043346} (\bibinfo {year} {2020})}\BibitemShut
  {NoStop}%
\bibitem [{\citenamefont {Biroli}\ and\ \citenamefont
  {Tarzia}(2021)}]{BirTar_Levy-RP}%
  \BibitemOpen
  \bibfield  {author} {\bibinfo {author} {\bibfnamefont {G.}~\bibnamefont
  {Biroli}}\ and\ \bibinfo {author} {\bibfnamefont {M.}~\bibnamefont
  {Tarzia}},\ }\bibfield  {title} {\bibinfo {title}
  {{L}\'evy-{R}osenzweig-{P}orter random matrix ensemble},\ }\href
  {https://doi.org/10.1103/PhysRevB.103.104205} {\bibfield  {journal} {\bibinfo
   {journal} {Phys. Rev. B}\ }\textbf {\bibinfo {volume} {103}},\ \bibinfo
  {pages} {104205} (\bibinfo {year} {2021})}\BibitemShut {NoStop}%
\bibitem [{\citenamefont {Khaymovich}\ and\ \citenamefont
  {Kravtsov}(2021)}]{LN-RP_dyn}%
  \BibitemOpen
  \bibfield  {author} {\bibinfo {author} {\bibfnamefont {I.~M.}\ \bibnamefont
  {Khaymovich}}\ and\ \bibinfo {author} {\bibfnamefont {V.~E.}\ \bibnamefont
  {Kravtsov}},\ }\bibfield  {title} {\bibinfo {title} {Dynamical phases in a
  ``multifractal'' {Rosenzweig}-{Porter} model},\ }\href
  {https://doi.org/10.21468/SciPostPhys.11.2.045} {\bibfield  {journal}
  {\bibinfo  {journal} {SciPost Phys.}\ }\textbf {\bibinfo {volume} {11}},\
  \bibinfo {pages} {45} (\bibinfo {year} {2021})}\BibitemShut {NoStop}%
\bibitem [{\citenamefont {Kutlin}\ and\ \citenamefont
  {Khaymovich}(2023)}]{Kutlin2023no_multifractal}%
  \BibitemOpen
  \bibfield  {author} {\bibinfo {author} {\bibfnamefont {A.~G.}\ \bibnamefont
  {Kutlin}}\ and\ \bibinfo {author} {\bibfnamefont {I.~M.}\ \bibnamefont
  {Khaymovich}},\ }\href
  {https://indico.fysik.su.se/event/7815/attachments/5052/6229/Kutlin_abstract.pdf}
  {\bibinfo {title} {Anatomy of the eigenstates distribution: a quest for a
  genuine multifractality}} (\bibinfo {year} {2023}),\ \bibinfo {note} {in
  preparation}\BibitemShut {NoStop}%
\bibitem [{\citenamefont {De~Tomasi}\ and\ \citenamefont
  {Khaymovich}(2022)}]{2022_nonHerm_RP}%
  \BibitemOpen
  \bibfield  {author} {\bibinfo {author} {\bibfnamefont {G.}~\bibnamefont
  {De~Tomasi}}\ and\ \bibinfo {author} {\bibfnamefont {I.~M.}\ \bibnamefont
  {Khaymovich}},\ }\bibfield  {title} {\bibinfo {title} {Non-hermitian
  rosenzweig-porter random-matrix ensemble: Obstruction to the fractal phase},\
  }\href {https://doi.org/10.1103/PhysRevB.106.094204} {\bibfield  {journal}
  {\bibinfo  {journal} {Phys. Rev. B}\ }\textbf {\bibinfo {volume} {106}},\
  \bibinfo {pages} {094204} (\bibinfo {year} {2022})}\BibitemShut {NoStop}%
\bibitem [{\citenamefont {Altshuler}\ and\ \citenamefont
  {Kravtsov}(2023)}]{Kravtsov2023Cantor}%
  \BibitemOpen
  \bibfield  {author} {\bibinfo {author} {\bibfnamefont {B.}~\bibnamefont
  {Altshuler}}\ and\ \bibinfo {author} {\bibfnamefont {V.}~\bibnamefont
  {Kravtsov}},\ }\bibfield  {title} {\bibinfo {title} {Random {Cantor} sets and
  mini-bands in local spectrum of quantum systems},\ }\href
  {https://doi.org/https://doi.org/10.1016/j.aop.2023.169300} {\bibfield
  {journal} {\bibinfo  {journal} {Annals of Physics}\ ,\ \bibinfo {pages}
  {169300}} (\bibinfo {year} {2023})},\ \bibinfo {note} {in press}\BibitemShut
  {NoStop}%
\bibitem [{Note1()}]{Note1}%
  \BibitemOpen
  \bibinfo {note} {Unlike the non-Hermitian case, where $1<d<2$ does change the
  phase diagram~\cite {2022_nonHerm_RP}.}\BibitemShut {Stop}%
\bibitem [{SM()}]{SM}%
  \BibitemOpen
  \href@noop {} {}\bibinfo {note} {See Supplemental Material at [URL will be
  inserted by publisher] for technical details, which includes
  Refs.~\cite{Biroli_RP,Monthus,BogomolnyRP2018,2022_nonHerm_RP,Kutlin2023no_multifractal}.}\BibitemShut
  {Stop}%
\bibitem [{Note2()}]{Note2}%
  \BibitemOpen
  \bibinfo {note} {Note that the strength of the diagonal elements is obtained
  from a fractal distribution, and not their spatial spread}\BibitemShut
  {NoStop}%
\bibitem [{\citenamefont {de~Tomasi}\ \emph {et~al.}(2019)\citenamefont
  {de~Tomasi}, \citenamefont {Amini}, \citenamefont {Bera}, \citenamefont
  {Khaymovich},\ and\ \citenamefont {Kravtsov}}]{RP_R(t)_2018}%
  \BibitemOpen
  \bibfield  {author} {\bibinfo {author} {\bibfnamefont {G.}~\bibnamefont
  {de~Tomasi}}, \bibinfo {author} {\bibfnamefont {M.}~\bibnamefont {Amini}},
  \bibinfo {author} {\bibfnamefont {S.}~\bibnamefont {Bera}}, \bibinfo {author}
  {\bibfnamefont {I.~M.}\ \bibnamefont {Khaymovich}},\ and\ \bibinfo {author}
  {\bibfnamefont {V.~E.}\ \bibnamefont {Kravtsov}},\ }\bibfield  {title}
  {\bibinfo {title} {Survival probability in generalized {Rosenzweig}-{Porter}
  random matrix ensemble},\ }\href
  {https://doi.org/10.21468/SciPostPhys.6.1.014} {\bibfield  {journal}
  {\bibinfo  {journal} {SciPost Phys.}\ }\textbf {\bibinfo {volume} {6}},\
  \bibinfo {pages} {014} (\bibinfo {year} {2019})}\BibitemShut {NoStop}%
\bibitem [{\citenamefont {Buijsman}\ and\ \citenamefont
  {Lev}(2022)}]{Buijsman2022circular}%
  \BibitemOpen
  \bibfield  {author} {\bibinfo {author} {\bibfnamefont {W.}~\bibnamefont
  {Buijsman}}\ and\ \bibinfo {author} {\bibfnamefont {Y.~B.}\ \bibnamefont
  {Lev}},\ }\bibfield  {title} {\bibinfo {title} {{Circular {Rosenzweig-Porter}
  random matrix ensemble}},\ }\href
  {https://doi.org/10.21468/SciPostPhys.12.3.082} {\bibfield  {journal}
  {\bibinfo  {journal} {SciPost Phys.}\ }\textbf {\bibinfo {volume} {12}},\
  \bibinfo {pages} {82} (\bibinfo {year} {2022})}\BibitemShut {NoStop}%
\bibitem [{\citenamefont {Dovgoshey}\ \emph {et~al.}(2006)\citenamefont
  {Dovgoshey}, \citenamefont {Martio}, \citenamefont {Ryazanov},\ and\
  \citenamefont {Vuorinen}}]{DOVGOSHEY20061}%
  \BibitemOpen
  \bibfield  {author} {\bibinfo {author} {\bibfnamefont {O.}~\bibnamefont
  {Dovgoshey}}, \bibinfo {author} {\bibfnamefont {O.}~\bibnamefont {Martio}},
  \bibinfo {author} {\bibfnamefont {V.}~\bibnamefont {Ryazanov}},\ and\
  \bibinfo {author} {\bibfnamefont {M.}~\bibnamefont {Vuorinen}},\ }\bibfield
  {title} {\bibinfo {title} {The {Cantor} function},\ }\href
  {https://doi.org/https://doi.org/10.1016/j.exmath.2005.05.002} {\bibfield
  {journal} {\bibinfo  {journal} {Expositiones Mathematicae}\ }\textbf
  {\bibinfo {volume} {24}},\ \bibinfo {pages} {1} (\bibinfo {year}
  {2006})}\BibitemShut {NoStop}%
\bibitem [{\citenamefont {Arnold}(1983)}]{Arnold_1983}%
  \BibitemOpen
  \bibfield  {author} {\bibinfo {author} {\bibfnamefont {B.~C.}\ \bibnamefont
  {Arnold}},\ }\href@noop {} {\emph {\bibinfo {title} {Pareto distributions}}}\
  (\bibinfo  {publisher} {International co-operative Publishing House},\
  \bibinfo {year} {1983})\BibitemShut {NoStop}%
\bibitem [{\citenamefont {Oganesyan}\ and\ \citenamefont
  {Huse}(2007)}]{oganesyan2007localization}%
  \BibitemOpen
  \bibfield  {author} {\bibinfo {author} {\bibfnamefont {V.}~\bibnamefont
  {Oganesyan}}\ and\ \bibinfo {author} {\bibfnamefont {D.~A.}\ \bibnamefont
  {Huse}},\ }\bibfield  {title} {\bibinfo {title} {Localization of interacting
  fermions at high temperature},\ }\href
  {https://doi.org/10.1103/PhysRevB.75.155111} {\bibfield  {journal} {\bibinfo
  {journal} {Phys. Rev. B}\ }\textbf {\bibinfo {volume} {75}},\ \bibinfo
  {pages} {155111} (\bibinfo {year} {2007})}\BibitemShut {NoStop}%
\bibitem [{\citenamefont {Atas}\ \emph {et~al.}(2013)\citenamefont {Atas},
  \citenamefont {Bogomolny}, \citenamefont {Giraud},\ and\ \citenamefont
  {Roux}}]{Atas2013distribution}%
  \BibitemOpen
  \bibfield  {author} {\bibinfo {author} {\bibfnamefont {Y.~Y.}\ \bibnamefont
  {Atas}}, \bibinfo {author} {\bibfnamefont {E.}~\bibnamefont {Bogomolny}},
  \bibinfo {author} {\bibfnamefont {O.}~\bibnamefont {Giraud}},\ and\ \bibinfo
  {author} {\bibfnamefont {G.}~\bibnamefont {Roux}},\ }\bibfield  {title}
  {\bibinfo {title} {Distribution of the ratio of consecutive level spacings in
  random matrix ensembles},\ }\href
  {https://doi.org/10.1103/PhysRevLett.110.084101} {\bibfield  {journal}
  {\bibinfo  {journal} {Phys. Rev. Lett.}\ }\textbf {\bibinfo {volume} {110}},\
  \bibinfo {pages} {084101} (\bibinfo {year} {2013})}\BibitemShut {NoStop}%
\bibitem [{\citenamefont {Mirlin}\ \emph {et~al.}(1996)\citenamefont {Mirlin},
  \citenamefont {Fyodorov}, \citenamefont {Dittes}, \citenamefont {Quezada},\
  and\ \citenamefont {Seligman}}]{MirFyod1996}%
  \BibitemOpen
  \bibfield  {author} {\bibinfo {author} {\bibfnamefont {A.~D.}\ \bibnamefont
  {Mirlin}}, \bibinfo {author} {\bibfnamefont {Y.~V.}\ \bibnamefont
  {Fyodorov}}, \bibinfo {author} {\bibfnamefont {F.-M.}\ \bibnamefont
  {Dittes}}, \bibinfo {author} {\bibfnamefont {J.}~\bibnamefont {Quezada}},\
  and\ \bibinfo {author} {\bibfnamefont {T.~H.}\ \bibnamefont {Seligman}},\
  }\bibfield  {title} {\bibinfo {title} {Transition from localized to extended
  eigenstates in the ensemble of power-law random banded matrices},\ }\href
  {https://doi.org/10.1103/PhysRevE.54.3221} {\bibfield  {journal} {\bibinfo
  {journal} {Phys. Rev. E}\ }\textbf {\bibinfo {volume} {54}},\ \bibinfo
  {pages} {3221} (\bibinfo {year} {1996})}\BibitemShut {NoStop}%
\bibitem [{\citenamefont {Burin}\ and\ \citenamefont
  {Maksimov}(1989)}]{Burin1989}%
  \BibitemOpen
  \bibfield  {author} {\bibinfo {author} {\bibfnamefont {A.~L.}\ \bibnamefont
  {Burin}}\ and\ \bibinfo {author} {\bibfnamefont {L.~A.}\ \bibnamefont
  {Maksimov}},\ }\bibfield  {title} {\bibinfo {title} {Localization and
  delocalization of particles in disordered lattice with tunneling amplitude
  with $r^{-3}$ decay},\ }\href
  {http://www.jetpletters.ac.ru/ps/1129/article_17116.shtml} {\bibfield
  {journal} {\bibinfo  {journal} {JETP Lett.}\ }\textbf {\bibinfo {volume}
  {50}},\ \bibinfo {pages} {338} (\bibinfo {year} {1989})}\BibitemShut
  {NoStop}%
\bibitem [{\citenamefont {Deng}\ \emph {et~al.}(2018)\citenamefont {Deng},
  \citenamefont {Kravtsov}, \citenamefont {Shlyapnikov},\ and\ \citenamefont
  {Santos}}]{Deng2018duality}%
  \BibitemOpen
  \bibfield  {author} {\bibinfo {author} {\bibfnamefont {X.}~\bibnamefont
  {Deng}}, \bibinfo {author} {\bibfnamefont {V.}~\bibnamefont {Kravtsov}},
  \bibinfo {author} {\bibfnamefont {G.}~\bibnamefont {Shlyapnikov}},\ and\
  \bibinfo {author} {\bibfnamefont {L.}~\bibnamefont {Santos}},\ }\bibfield
  {title} {\bibinfo {title} {Duality in power-law localization in disordered
  one-dimensional systems},\ }\href
  {https://doi.org/10.1103/PhysRevLett.120.110602} {\bibfield  {journal}
  {\bibinfo  {journal} {Phys. Rev. Lett.}\ }\textbf {\bibinfo {volume} {120}},\
  \bibinfo {pages} {110602} (\bibinfo {year} {2018})}\BibitemShut {NoStop}%
\bibitem [{\citenamefont {Nosov}\ \emph {et~al.}(2019)\citenamefont {Nosov},
  \citenamefont {Khaymovich},\ and\ \citenamefont
  {Kravtsov}}]{Nosov2019correlation}%
  \BibitemOpen
  \bibfield  {author} {\bibinfo {author} {\bibfnamefont {P.~A.}\ \bibnamefont
  {Nosov}}, \bibinfo {author} {\bibfnamefont {I.~M.}\ \bibnamefont
  {Khaymovich}},\ and\ \bibinfo {author} {\bibfnamefont {V.~E.}\ \bibnamefont
  {Kravtsov}},\ }\bibfield  {title} {\bibinfo {title} {Correlation-induced
  localization},\ }\href {https://doi.org/10.1103/PhysRevB.99.104203}
  {\bibfield  {journal} {\bibinfo  {journal} {Physical Review B}\ }\textbf
  {\bibinfo {volume} {99}},\ \bibinfo {pages} {104203} (\bibinfo {year}
  {2019})}\BibitemShut {NoStop}%
\bibitem [{\citenamefont {Deng}\ \emph {et~al.}(2022)\citenamefont {Deng},
  \citenamefont {Burin},\ and\ \citenamefont
  {Khaymovich}}]{deng2020anisotropy}%
  \BibitemOpen
  \bibfield  {author} {\bibinfo {author} {\bibfnamefont {X.}~\bibnamefont
  {Deng}}, \bibinfo {author} {\bibfnamefont {A.~L.}\ \bibnamefont {Burin}},\
  and\ \bibinfo {author} {\bibfnamefont {I.~M.}\ \bibnamefont {Khaymovich}},\
  }\bibfield  {title} {\bibinfo {title} {{Anisotropy-mediated reentrant
  localization}},\ }\href {https://doi.org/10.21468/SciPostPhys.13.5.116}
  {\bibfield  {journal} {\bibinfo  {journal} {SciPost Phys.}\ }\textbf
  {\bibinfo {volume} {13}},\ \bibinfo {pages} {116} (\bibinfo {year}
  {2022})}\BibitemShut {NoStop}%
\bibitem [{\citenamefont {Richardson}(1963)}]{Richardson1963restricted}%
  \BibitemOpen
  \bibfield  {author} {\bibinfo {author} {\bibfnamefont {R.}~\bibnamefont
  {Richardson}},\ }\bibfield  {title} {\bibinfo {title} {A restricted class of
  exact eigenstates of the pairing-force {Hamiltonian}},\ }\href
  {https://doi.org/10.1016/0031-9163(63)90259-2} {\bibfield  {journal}
  {\bibinfo  {journal} {Phys. Lett.}\ }\textbf {\bibinfo {volume} {3}},\
  \bibinfo {pages} {277} (\bibinfo {year} {1963})}\BibitemShut {NoStop}%
\bibitem [{\citenamefont {Richardson}\ and\ \citenamefont
  {Sherman}(1964)}]{Richardson1964exact}%
  \BibitemOpen
  \bibfield  {author} {\bibinfo {author} {\bibfnamefont {R.}~\bibnamefont
  {Richardson}}\ and\ \bibinfo {author} {\bibfnamefont {N.}~\bibnamefont
  {Sherman}},\ }\bibfield  {title} {\bibinfo {title} {Exact eigenstates of the
  pairing-force {Hamiltonian}},\ }\href
  {https://doi.org/10.1016/0029-5582(64)90687-X} {\bibfield  {journal}
  {\bibinfo  {journal} {Nuclear Physics}\ }\textbf {\bibinfo {volume} {52}},\
  \bibinfo {pages} {221} (\bibinfo {year} {1964})}\BibitemShut {NoStop}%
\bibitem [{\citenamefont {Modak}\ \emph {et~al.}(2016)\citenamefont {Modak},
  \citenamefont {Mukerjee}, \citenamefont {Yuzbashyan},\ and\ \citenamefont
  {Shastry}}]{Modak2016integrals}%
  \BibitemOpen
  \bibfield  {author} {\bibinfo {author} {\bibfnamefont {R.}~\bibnamefont
  {Modak}}, \bibinfo {author} {\bibfnamefont {S.}~\bibnamefont {Mukerjee}},
  \bibinfo {author} {\bibfnamefont {E.~A.}\ \bibnamefont {Yuzbashyan}},\ and\
  \bibinfo {author} {\bibfnamefont {B.~S.}\ \bibnamefont {Shastry}},\
  }\bibfield  {title} {\bibinfo {title} {Integrals of motion for
  one-dimensional {Anderson} localized systems},\ }\href
  {https://doi.org/10.1088/1367-2630/18/3/033010} {\bibfield  {journal}
  {\bibinfo  {journal} {New J. Phys.}\ }\textbf {\bibinfo {volume} {18}},\
  \bibinfo {pages} {033010} (\bibinfo {year} {2016})}\BibitemShut {NoStop}%
\bibitem [{\citenamefont {V.E.Kravtsov}\ \emph {et~al.}(2018)\citenamefont
  {V.E.Kravtsov}, \citenamefont {B.L.Altshuler},\ and\ \citenamefont
  {L.B.Ioffe}}]{RRGAnnals}%
  \BibitemOpen
  \bibfield  {author} {\bibinfo {author} {\bibnamefont {V.E.Kravtsov}},
  \bibinfo {author} {\bibnamefont {B.L.Altshuler}},\ and\ \bibinfo {author}
  {\bibnamefont {L.B.Ioffe}},\ }\bibfield  {title} {\bibinfo {title}
  {Non-ergodic delocalized phase in {Anderson} model on {Bethe} lattice and
  regular graph},\ }\href
  {https://doi.org/https://doi.org/10.1016/j.aop.2017.12.009} {\bibfield
  {journal} {\bibinfo  {journal} {Annals of Physics}\ }\textbf {\bibinfo
  {volume} {389}},\ \bibinfo {pages} {148} (\bibinfo {year}
  {2018})}\BibitemShut {NoStop}%
\bibitem [{\citenamefont {Tikhonov}\ and\ \citenamefont
  {Mirlin}(2021)}]{Tikh-Mir_RRG_review}%
  \BibitemOpen
  \bibfield  {author} {\bibinfo {author} {\bibfnamefont {K.}~\bibnamefont
  {Tikhonov}}\ and\ \bibinfo {author} {\bibfnamefont {A.}~\bibnamefont
  {Mirlin}},\ }\bibfield  {title} {\bibinfo {title} {From {Anderson}
  localization on random regular graphs to many-body localization},\ }\href
  {https://doi.org/https://doi.org/10.1016/j.aop.2021.168525} {\bibfield
  {journal} {\bibinfo  {journal} {Annals of Physics}\ ,\ \bibinfo {pages}
  {168525}} (\bibinfo {year} {2021})}\BibitemShut {NoStop}%
\bibitem [{\citenamefont {Smelyanskiy}\ \emph {et~al.}(2020)\citenamefont
  {Smelyanskiy}, \citenamefont {Kechedzhi}, \citenamefont {Boixo},
  \citenamefont {Isakov}, \citenamefont {Neven},\ and\ \citenamefont
  {Altshuler}}]{Smelyansky_Grover}%
  \BibitemOpen
  \bibfield  {author} {\bibinfo {author} {\bibfnamefont {V.~N.}\ \bibnamefont
  {Smelyanskiy}}, \bibinfo {author} {\bibfnamefont {K.}~\bibnamefont
  {Kechedzhi}}, \bibinfo {author} {\bibfnamefont {S.}~\bibnamefont {Boixo}},
  \bibinfo {author} {\bibfnamefont {S.~V.}\ \bibnamefont {Isakov}}, \bibinfo
  {author} {\bibfnamefont {H.}~\bibnamefont {Neven}},\ and\ \bibinfo {author}
  {\bibfnamefont {B.}~\bibnamefont {Altshuler}},\ }\bibfield  {title} {\bibinfo
  {title} {Nonergodic delocalized states for efficient population transfer
  within a narrow band of the energy landscape},\ }\href
  {https://doi.org/10.1103/PhysRevX.10.011017} {\bibfield  {journal} {\bibinfo
  {journal} {Phys. Rev. X}\ }\textbf {\bibinfo {volume} {10}},\ \bibinfo
  {pages} {011017} (\bibinfo {year} {2020})}\BibitemShut {NoStop}%
\bibitem [{\citenamefont {Kechedzhi}\ \emph {et~al.}(2018)\citenamefont
  {Kechedzhi}, \citenamefont {Smelyanskiy}, \citenamefont {McClean},
  \citenamefont {Denchev}, \citenamefont {Mohseni}, \citenamefont {Isakov},
  \citenamefont {Boixo}, \citenamefont {Altshuler},\ and\ \citenamefont
  {Neven}}]{Smelyansky_ML}%
  \BibitemOpen
  \bibfield  {author} {\bibinfo {author} {\bibfnamefont {K.}~\bibnamefont
  {Kechedzhi}}, \bibinfo {author} {\bibfnamefont {V.~N.}\ \bibnamefont
  {Smelyanskiy}}, \bibinfo {author} {\bibfnamefont {J.~R.}\ \bibnamefont
  {McClean}}, \bibinfo {author} {\bibfnamefont {V.~S.}\ \bibnamefont
  {Denchev}}, \bibinfo {author} {\bibfnamefont {M.}~\bibnamefont {Mohseni}},
  \bibinfo {author} {\bibfnamefont {S.~V.}\ \bibnamefont {Isakov}}, \bibinfo
  {author} {\bibfnamefont {S.}~\bibnamefont {Boixo}}, \bibinfo {author}
  {\bibfnamefont {B.~L.}\ \bibnamefont {Altshuler}},\ and\ \bibinfo {author}
  {\bibfnamefont {H.}~\bibnamefont {Neven}},\ }\bibfield  {title} {\bibinfo
  {title} {Efficient population transfer via non-ergodic extended states in
  quantum spin glass},\ }\Eprint {https://arxiv.org/abs/1807.04792}
  {arXiv:1807.04792}  (\bibinfo {year} {2018})\BibitemShut {NoStop}%
\end{thebibliography}%


\begin{thebibliography}{8}%
\makeatletter
\providecommand \@ifxundefined [1]{%
 \@ifx{#1\undefined}
}%
\providecommand \@ifnum [1]{%
 \ifnum #1\expandafter \@firstoftwo
 \else \expandafter \@secondoftwo
 \fi
}%
\providecommand \@ifx [1]{%
 \ifx #1\expandafter \@firstoftwo
 \else \expandafter \@secondoftwo
 \fi
}%
\providecommand \natexlab [1]{#1}%
\providecommand \enquote  [1]{``#1''}%
\providecommand \bibnamefont  [1]{#1}%
\providecommand \bibfnamefont [1]{#1}%
\providecommand \citenamefont [1]{#1}%
\providecommand \href@noop [0]{\@secondoftwo}%
\providecommand \href [0]{\begingroup \@sanitize@url \@href}%
\providecommand \@href[1]{\@@startlink{#1}\@@href}%
\providecommand \@@href[1]{\endgroup#1\@@endlink}%
\providecommand \@sanitize@url [0]{\catcode `\\12\catcode `\$12\catcode
  `\&12\catcode `\#12\catcode `\^12\catcode `\_12\catcode `\%12\relax}%
\providecommand \@@startlink[1]{}%
\providecommand \@@endlink[0]{}%
\providecommand \url  [0]{\begingroup\@sanitize@url \@url }%
\providecommand \@url [1]{\endgroup\@href {#1}{\urlprefix }}%
\providecommand \urlprefix  [0]{URL }%
\providecommand \Eprint [0]{\href }%
\providecommand \doibase [0]{https://doi.org/}%
\providecommand \selectlanguage [0]{\@gobble}%
\providecommand \bibinfo  [0]{\@secondoftwo}%
\providecommand \bibfield  [0]{\@secondoftwo}%
\providecommand \translation [1]{[#1]}%
\providecommand \BibitemOpen [0]{}%
\providecommand \bibitemStop [0]{}%
\providecommand \bibitemNoStop [0]{.\EOS\space}%
\providecommand \EOS [0]{\spacefactor3000\relax}%
\providecommand \BibitemShut  [1]{\csname bibitem#1\endcsname}%
\let\auto@bib@innerbib\@empty
\bibitem [{\citenamefont {Facoetti}\ \emph {et~al.}(2016)\citenamefont
  {Facoetti}, \citenamefont {Vivo},\ and\ \citenamefont {Biroli}}]{Biroli_RP}%
  \BibitemOpen
  \bibfield  {author} {\bibinfo {author} {\bibfnamefont {D.}~\bibnamefont
  {Facoetti}}, \bibinfo {author} {\bibfnamefont {P.}~\bibnamefont {Vivo}},\
  and\ \bibinfo {author} {\bibfnamefont {G.}~\bibnamefont {Biroli}},\
  }\bibfield  {title} {\bibinfo {title} {From non-ergodic eigenvectors to local
  resolvent statistics and back: A random matrix perspective},\ }\href
  {https://doi.org/10.1209/0295-5075/115/47003} {\bibfield  {journal} {\bibinfo
   {journal} {Europhys. Lett.}\ }\textbf {\bibinfo {volume} {115}},\ \bibinfo
  {pages} {47003} (\bibinfo {year} {2016})}\BibitemShut {NoStop}%
\bibitem [{\citenamefont {Monthus}(2017)}]{Monthus}%
  \BibitemOpen
  \bibfield  {author} {\bibinfo {author} {\bibfnamefont {C.}~\bibnamefont
  {Monthus}},\ }\bibfield  {title} {\bibinfo {title} {Statistical properties of
  the {Green} function in finite size for {Anderson} localization models with
  multifractal eigenvectors},\ }\href
  {https://doi.org/10.1088/1751-8121/aa5ad2} {\bibfield  {journal} {\bibinfo
  {journal} {J. Phys. A: Math. Theor.}\ }\textbf {\bibinfo {volume} {50}},\
  \bibinfo {pages} {295101} (\bibinfo {year} {2017})}\BibitemShut {NoStop}%
\bibitem [{\citenamefont {Bogomolny}\ and\ \citenamefont
  {Sieber}(2018)}]{BogomolnyRP2018}%
  \BibitemOpen
  \bibfield  {author} {\bibinfo {author} {\bibfnamefont {E.}~\bibnamefont
  {Bogomolny}}\ and\ \bibinfo {author} {\bibfnamefont {M.}~\bibnamefont
  {Sieber}},\ }\bibfield  {title} {\bibinfo {title} {Eigenfunction distribution
  for the {Rosenzweig}-{Porter} model},\ }\href
  {https://doi.org/10.1103/PhysRevE.98.032139} {\bibfield  {journal} {\bibinfo
  {journal} {Phys. Rev. E}\ }\textbf {\bibinfo {volume} {98}},\ \bibinfo
  {pages} {032139} (\bibinfo {year} {2018})}\BibitemShut {NoStop}%
\bibitem [{\citenamefont {De~Tomasi}\ and\ \citenamefont
  {Khaymovich}(2022)}]{2022_nonHerm_RP}%
  \BibitemOpen
  \bibfield  {author} {\bibinfo {author} {\bibfnamefont {G.}~\bibnamefont
  {De~Tomasi}}\ and\ \bibinfo {author} {\bibfnamefont {I.~M.}\ \bibnamefont
  {Khaymovich}},\ }\bibfield  {title} {\bibinfo {title} {Non-hermitian
  rosenzweig-porter random-matrix ensemble: Obstruction to the fractal phase},\
  }\href {https://doi.org/10.1103/PhysRevB.106.094204} {\bibfield  {journal}
  {\bibinfo  {journal} {Phys. Rev. B}\ }\textbf {\bibinfo {volume} {106}},\
  \bibinfo {pages} {094204} (\bibinfo {year} {2022})}\BibitemShut {NoStop}%
\bibitem [{\citenamefont {Kutlin}\ and\ \citenamefont
  {Khaymovich}(2021)}]{Kutlin2021emergent}%
  \BibitemOpen
  \bibfield  {author} {\bibinfo {author} {\bibfnamefont {A.~G.}\ \bibnamefont
  {Kutlin}}\ and\ \bibinfo {author} {\bibfnamefont {I.~M.}\ \bibnamefont
  {Khaymovich}},\ }\bibfield  {title} {\bibinfo {title} {Emergent fractal phase
  in energy stratified random models},\ }\href
  {https://doi.org/10.21468/SciPostPhys.11.6.101} {\bibfield  {journal}
  {\bibinfo  {journal} {SciPost Phys.}\ }\textbf {\bibinfo {volume} {11}},\
  \bibinfo {pages} {101} (\bibinfo {year} {2021})}\BibitemShut {NoStop}%
\bibitem [{\citenamefont {Tarzia}(2020)}]{Tarzia_2020}%
  \BibitemOpen
  \bibfield  {author} {\bibinfo {author} {\bibfnamefont {M.}~\bibnamefont
  {Tarzia}},\ }\bibfield  {title} {\bibinfo {title} {Many-body localization
  transition in {Hilbert space}},\ }\href
  {https://doi.org/10.1103/PhysRevB.102.014208} {\bibfield  {journal} {\bibinfo
   {journal} {Phys. Rev. B}\ }\textbf {\bibinfo {volume} {102}},\ \bibinfo
  {pages} {014208} (\bibinfo {year} {2020})}\BibitemShut {NoStop}%
\bibitem [{\citenamefont {Kutlin}\ and\ \citenamefont
  {Khaymovich}(2023)}]{Kutlin2023no_multifractal}%
  \BibitemOpen
  \bibfield  {author} {\bibinfo {author} {\bibfnamefont {A.~G.}\ \bibnamefont
  {Kutlin}}\ and\ \bibinfo {author} {\bibfnamefont {I.~M.}\ \bibnamefont
  {Khaymovich}},\ }\href
  {https://indico.fysik.su.se/event/7815/attachments/5052/6229/Kutlin_abstract.pdf}
  {\bibinfo {title} {Anatomy of the eigenstates distribution: a quest for a
  genuine multifractality}} (\bibinfo {year} {2023}),\ \bibinfo {note} {in
  preparation}\BibitemShut {NoStop}%
\bibitem [{\citenamefont {de~Tomasi}\ \emph {et~al.}(2019)\citenamefont
  {de~Tomasi}, \citenamefont {Amini}, \citenamefont {Bera}, \citenamefont
  {Khaymovich},\ and\ \citenamefont {Kravtsov}}]{RP_R(t)_2018}%
  \BibitemOpen
  \bibfield  {author} {\bibinfo {author} {\bibfnamefont {G.}~\bibnamefont
  {de~Tomasi}}, \bibinfo {author} {\bibfnamefont {M.}~\bibnamefont {Amini}},
  \bibinfo {author} {\bibfnamefont {S.}~\bibnamefont {Bera}}, \bibinfo {author}
  {\bibfnamefont {I.~M.}\ \bibnamefont {Khaymovich}},\ and\ \bibinfo {author}
  {\bibfnamefont {V.~E.}\ \bibnamefont {Kravtsov}},\ }\bibfield  {title}
  {\bibinfo {title} {Survival probability in generalized {Rosenzweig}-{Porter}
  random matrix ensemble},\ }\href
  {https://doi.org/10.21468/SciPostPhys.6.1.014} {\bibfield  {journal}
  {\bibinfo  {journal} {SciPost Phys.}\ }\textbf {\bibinfo {volume} {6}},\
  \bibinfo {pages} {014} (\bibinfo {year} {2019})}\BibitemShut {NoStop}%
\end{thebibliography}%

\end{document}


\title{Supplementary material for ``Tuning the phase diagram of a Rosenzweig-Porter model with fractal disorder"}
\author{Madhumita Sarkar }
\thanks{sarkar.madhumita770@gmail.com}
\affiliation{Jo{\v z}ef Stefan Institute, SI-1000, Ljubljana, Slovenia}
\author{Roopayan Ghosh}
\affiliation{Department of Physics and Astronomy, University College London, Gower Street, WC1E6BT, London}
\author{Ivan Khaymovich}
\affiliation{Nordita, Stockholm University and KTH Royal Institute of
Technology Hannes Alfv{\'e}ns v{\"a}g 12, SE-106 91 Stockholm, Sweden}
\affiliation{Institute for Physics of Microstructures, Russian Academy of Sciences, 603950 Nizhny Novgorod, GSP-105, Russia}
\maketitle
\section{Derivation of Eq.~(6) and Eq.~(7)}
\label{app:appA}
The cavity Green's function method is a technique to self consistently solve the Green's function of a system by connecting the complete Green's function of the system to the Green's function with a single site removed. 
Green's function is defined as
\begin{equation}
 G (E+i \delta)= (E+i \delta -H)^{-1} \ .
\end{equation}
\imkR{Here $\eta$ is an infinitisemal regularizer.}
Using the Schur's complement formula at the removed site $m$, the cavity equation takes the form~\cite{Biroli_RP,Monthus, BogomolnyRP2018}
\begin{multline}
 G_{mm} (E+i \delta)= \lrp{E+i \delta-h_m +\sum_{n, r \neq m}H_{mn} G^{ (m)}_{nr}H_{rm}}^{-1}
 \label{eq:Green}
\end{multline}
where $G^{ (m)}$ denotes the Green's function with the $m^{th}$ row and column removed, $H_{mn} = L^{-\gamma/2}M_{mn}$. 

As the self-energy $\Sigma=\sum_{n, r \neq m}H_{mn} G^{ (m)}_{nr}H_{rm}$ is self-averaging for not fat-tail distributed $H_{m\ne n}$, following the literature~\cite{Biroli_RP,BogomolnyRP2018,2022_nonHerm_RP}, one can consider only diagonal part of it, which is not averaged out
\begin{align}
 \Sigma\simeq \overline{\Sigma}=\sum_n \overline{H_{mn}^2}  \overline{G_{nn}^{(m)}}= 
 L^{-\gamma} \sum_n G_{nn} \ .
\end{align}
In the last equality we have replaced $G^{(m)}$ by $G$, which is exact in the thermodynamic limit, and substituted the variance of $H_{m\ne n}$.

Substituting Eq.~\eqref{eq:Green} into the latter, one obtains the self-consistency equation
\be
\overline{\Sigma} = \sum_n \frac{L^{-\gamma}}{E-h_n +i \delta+ \overline{\Sigma}}
\ee
which is equivalent to Eq.(6) for $\overline\Gamma = \Im \overline \Sigma$ and neglected $\Re \overline \Sigma \ll E, \ep_i$. \imkR{This self-energy $\overline\Gamma$ of a Green's function corresponds to the wave-function fractal support set $\overline\Gamma\sim L^{D-1}$ as soon as one can consider regularizer $\eta$ large compared to the typical level spacing $\eta\gg \delta_{typ}$~\cite{Kutlin2021emergent,Tarzia_2020}. Indeed, in this case the broadening itself is large $\overline \Gamma\gg \delta_{typ}$ and the self-averaging property as well as the saddle-point approximation below are valid.}

Then without loss of generality, we can rewrite Eq.~(6) as,
\begin{align}\label{eq:self_consist_eq}
1 &= \sum_n \frac{L^{-\gamma}}{|E -h_n|^2 +\bar \Gamma^2}\nonumber \\ 
&=\sum_b L^{1-f (b)} \frac{L^{-\gamma}}{L^{-2b} +L^{-2a}} \nonumber \\
&\simeq
\sum_{b<a} L^{1-\gamma-f (b)+2b} + \sum_{b>a} L^{1-\gamma-f (b)+2a}
\end{align}
In the second step we have changed the summation index from $n$ to $b$, using the fact that the number of levels with $|E-h_n|\sim L^{-b}$ is $L^{1-f (b)}$, and in the third step we approximated the denominator by the larger of $L^{-2b}$
 and $L^{-2a}$ at the appropriate values of $b$, since that is the dominant term as $L \rightarrow \infty$.

To evaluate the expression, notice that the function in the first part is monotonically increasing with $b$, while in the second one it is monotonically decreasing with $b$. Then, within the saddle-point approximation (ignoring the pre-factors), the sum is given by the maximal value of the summand at $b=a$. Doing so we arrive at Eq.~(7) in the main text.

\section{Multifractal disorder}
\label{app:appB}
In order to realize a multifractal disorder, following the same idea as in Eqs.~(14),~(15), one should take a generic multifractal probability distribution $P (s)$ (written in a saddle-point approximation)
\be\label{eq:P (s)_g (nu)}
\begin{aligned}
&P\lrp{s\sim N^{-\nu}}ds \sim N^{g (\nu)-1}\mathbf{d}\nu \ , \\
&\quad \max_\nu g (\nu) \equiv g (\nu_0) = 1 \ , \\
&\quad \max_\nu \lrb{g (\nu) - q\nu}\equiv g (\nu_q)-q\cdot \nu_q \ ,
\end{aligned}
\ee
where we introduced the notations $\nu_q$ for the moments
\be
\mean{s^q}\sim N^{g (\nu_q)-1-q\cdot \nu_q} \ ,
\ee
to which the main contribution is given by $s_q \sim N^{-\nu_q}$.
$\delta_{typ} \sim N^{-\nu_0}$ is the typical (most probable) value of $s$, while the mean-level spacing, given by the first moment (if it converges)
\be\label{eq:delta}
\delta = \mean{s} \sim N^{g (\nu_1)-\nu_1-1}\sim 1/N \lra g (\nu_1) = \nu_1 \ .
\ee
In the last equality we assume a bandwidth $E_{BW}\equiv N\delta\sim N^0$ to be finite.

Note that for a smooth function $g (\nu)$ the Legendre transform in Eq.~\eqref{eq:P (s)_g (nu)} gives the condition
\be\label{eq:g' (nu_q)}
g' (\nu_q) = q 
\ee
while 
the prefactor in Eq.~\eqref{eq:P (s)_g (nu)} within a saddle-point approximation is given by
\be\label{eq:P (s)_g (nu)_prefactor2}
P\lrp{s\sim N^{-\nu}}\mathbf{d}s = \sqrt{\frac{|g'' (\nu_0)|\ln N}{2\pi}} N^{g (\nu)-1}\mathbf{d}\nu \ ,
\ee
where the smoothness of $g(\nu)$ at $\nu=\nu_0$ guarantees finiteness of the second derivative. 

\paragraph{Correspondence between fractality of spectrum $f(b)$ and level spacing distribution $g(\nu)$.}

In order to understand how the levels $\zeta_n$ are distributed according to Eq.~(3), one should take the extensive number of them
$m \equiv N^{1-f (b)}$ and calculate the distance between
\be\label{eq:sum_sk}
\zeta_{n+m}-\zeta_{n}\equiv \sum_{k=1}^{m}{s_{n+k}} \sim N^{-b} \ ,
\ee
which gives an estimate of the number of levels in between.
Please note that here, unlike Eq.~(3), our control parameter will be the scaling of $m$, but not $N^{-b}$.

In order to find the correspondence between $f (b)$ and $g (\nu)$ we will use the method, developed in~\cite{Kutlin2023no_multifractal} to find a distribution of the extensive sums of multifractal i.i.d. random numbers.
Indeed, for any $f (b)$ (or $m$) there can be two types of main contributions to Eq.~\eqref{eq:sum_sk}: the individual one and the collective one.
In order to understand it, we will separate the sum of $m$ elements into the ``bins'', close to a certain $\nu$, $s\in[N^{-\nu-\mathbf{d}\nu},N^{-\nu})$.
For each of these bins, the number of elements in the corresponding part of the sum is given by
\be\label{eq:Num_nu}
M_\nu\sim m\cdot P (s)ds\sim N^{g (\nu)-f (b)}\mathbf{d}\nu \ .
\ee
Within the saddle-point approximation, the parameter $b$ is determined by the maximum of the collective and individual contributions.

The individual contribution is given by a maximal $s_k \sim N^{-\nu^*}$ which appears in the above sum Eq.~\eqref{eq:sum_sk} at least once, $M_{\nu^*} \sim N^0$.
Thus, $\nu^*$ is given by the smaller solution of the equation
\be
g (\nu^*) = f (b) \ , \nu^* = b \ .
\ee
Thus, the amplitude of the sum is at least given by $N^{-b}\gtrsim N^{-\nu^*}$.
The saturation of the latter inequality, $b\leq \nu^*$, gives $f (b) = g (b)$.

The collective contribution means that the corresponding number $M_\nu$ of elements $s_k\sim N^{-\nu}$, Eq.~\eqref{eq:Num_nu}, is extensive there, i.e., $g(\nu)>f(b)$.
Similarly to Eq.~\eqref{eq:self_consist_eq} changing the summation from $k$ to $\nu$ in Eq.~\eqref{eq:sum_sk} we obtain
\be
N^{-b} \sim m\mean{s}_{M_{\nu}\gg 1} \sim \int_{g (\nu)>f (b)} \mathbf{d}\nu N^{g (\nu)-f (b)-\nu} \ .
\ee
The latter integral for $g (\nu_1)>f (b)$ is given by $\nu=\nu_1$, i.e. by the mean-level spacing Eq.~\eqref{eq:delta}.
This gives $b = f (b)$ as $N\delta \sim N^{g (\nu_1)-\nu_1}\sim N^0$.
Otherwise, $g (\nu_1)<f (b)$, the integral sits at $g (\nu^*)=f (b)$ and is dominated by the individual contribution.

Summarizing both cases, one obtains
\be
N^{-b} \sim \left\{
N^{-\nu^*} \ , \; g (\nu^*)=f (b) \ , \; g (\nu_1)<f (b) \atop
N^{g (\nu_1)-f (b)-\nu_1} \ , \quad\quad g (\nu_1)>f (b)
\right.
\ee
This result is quite straightforward as we know from the properties of $f (b)$ that its derivative cannot be larger than $1$. The latter case corresponds to $d=1$ in the fractal regime Eq.~(4).

Hence, correspondence between $f (b)$ and $g (\nu)$ can be found to be
\beq
f (b) = 
\left\{
b \ , \quad b<\nu_1 \atop
g (b) \ , \quad b>\nu_1
\right. 
\eeq

\paragraph{Log normal distribution as an example}
Let's consider a log-normal distribution of $s$, then $P (\nu)$ should be Gaussian, i.e., $g (\nu)$ is a parabola
\be
g (\nu) = 1 - A (\nu-\nu_0)^2 \ ,
\ee
where we have parameterized it with the location of the maximum (typical $s_{typ}\sim N^{-\nu_0}$) and used the normalization condition $g (\nu_0)=1$ from Eq.~\eqref{eq:P (s)_g (nu)}.
For this choice the prefactor in Eq.~(16) is exact.

Taking into account also Eqs.~\eqref{eq:delta} and~\eqref{eq:g' (nu_q)}, one obtains
\be
\label{eq:lognormal}
\begin{aligned}
1 &= g' (\nu_1) = - 2 A (\nu_1-\nu_0) \lra A = \frac{1}{2 (\nu_0-\nu_1)}>0 \ ,\\
\nu_1 &= g (\nu_1) = 1 + \frac{\nu_1-\nu_0}{2} \lra \nu_1 = 2-\nu_0 < 1 \ ,\\
g (\nu) &= 1 - \frac{ (\nu-\nu_0)^2}{4 (\nu_0-1)} \ .
\end{aligned}
\ee


Next, using Eqs.~(7) and~(8), for the log-normal distribution in Eq.~\eqref{eq:lognormal} one obtains the following equations for $b$

\be
b = \gamma-1 \ , 
\ee
at $\gamma<3-\nu_0$ and
\be
b^2+2b (3\nu_0-4)+ \nu_0^2-4 (\nu_0-1)\gamma = 0 
\ee
at $\gamma>3-\nu_0$, with the solution
\be
b = \left\{
\begin{array}{cl}
\gamma-1 \ , & \gamma<3-\nu_0 \\
4-3\nu_0 +2\sqrt{ (\nu_0-1) (\gamma+2\nu_0 - 4)} \ , & \gamma>3-\nu_0
\end{array}
\right. 
\ee
leading to Eq.~(17) in the main text.
\section{Further data about $r$ statistics}
\label{app:appC}
\begin{figure}
\centering {\ing[width=0.9\linewidth]
{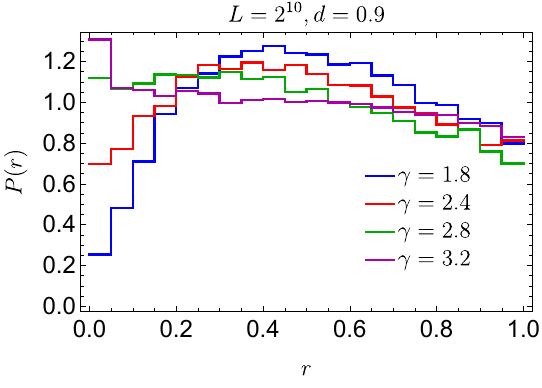}}
\caption{Histogram of Probability density function of consecutive level spacing ratio $r$ for diagonal disorder having Hausdorff dimension $d=0.9$ for different $\gamma=1.8,2.4,2.8,3.2$ chosen near the ``kink" in Fig.~5(d)}
\label{fig:rstat2}
\end{figure}
In Fig.~\ref{fig:rstat2} we show the histogram of probability density function of $r$ defined in Eq.~(18) for different $\gamma$ to shed some light on the source of the second kink. From Fig.~5, we see that the second kink occurs inside the localized phase (for $d=0.9$ at $\gamma>\gamma_{AT}\sim 2.2$). Looking at Fig.~\ref{fig:rstat2} we can verify that indeed already at $\gamma=2.4$, the level repulsion is sufficiently weak. However the distribution undergoes further changes as we increase $\gamma$ not only in the regime of $r \sim 0$, where we have increasing weight, but also in the region $r \sim 1$ due our choice of diagonal disorder. Deep inside the localized phase the energy levels are completely non-hybridized and the eigenenergies would be given by unperturbed diagonal elements. Since $P(r) \sim 1/r^{1-d}$, for $d\sim 1$ this indicates large weight at $r \sim 1$ which is seen for the $\gamma=3.2$ line in the plot, which has a larger weight in that regime than $\gamma \sim 2.4,2.8$. Hence the mean, $\langle r \rangle$ shows a rise at large values of $\gamma$ for $d \sim 1$.

\section{Survival probability:}
\label{app:AppD}
We conclude our analysis by studying the dynamical behaviour of this model. Specifically, we consider the time dependence of the survival (or return) probability, $R (t)$~\cite{RP_R(t)_2018}, defined by,
\be
R (t)=|\langle \psi (0)| e^{-i H t}|\psi (0) \rangle|^2,
\ee
in various regimes. We consider the initial wave function, localized on a single site, $N$, i.e. $|\psi (0)\rangle=\sum_{i=1}^L\delta_{i,N}\langle i|\psi (0)\rangle$.

Like in the standard RP-model~\cite{RP_R(t)_2018}, in the NEE phase $\mathcal{R} (t) \sim e^{-\Gamma t}$, where $\Gamma= L^{-\frac{\gamma -1}{2-d}}$, Eq.~(10).
\begin{figure}
\centering {\ing[width=0.48\linewidth,height=0.36\columnwidth]{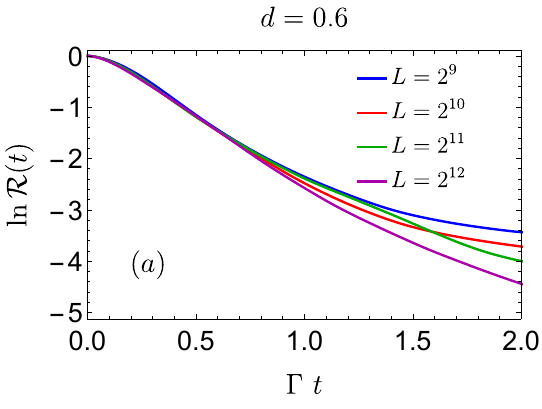}}
\centering {\ing[width=0.48\linewidth,height=0.36\columnwidth]{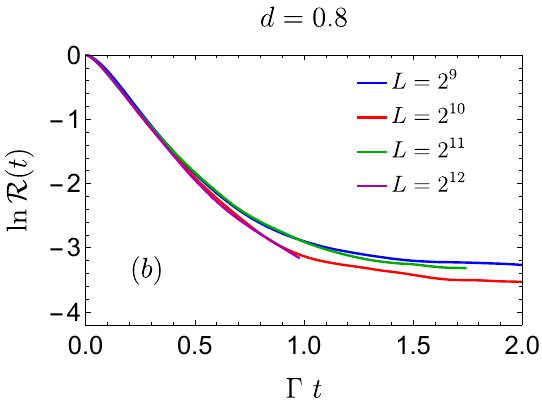}}
\centering {\ing[width=0.48\linewidth,height=0.36\columnwidth]{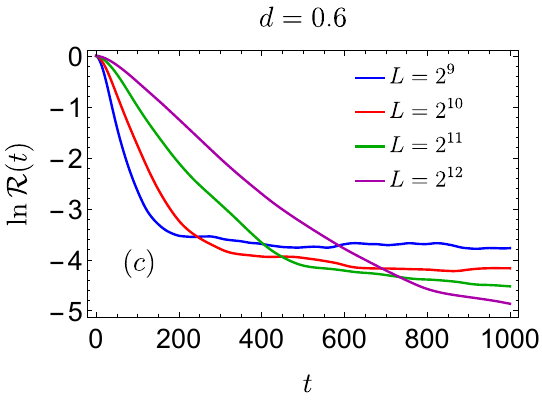}}
\centering {\ing[width=0.48\linewidth,height=0.36\columnwidth]{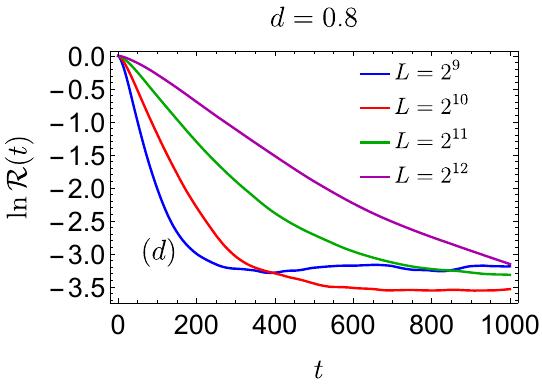}}
\caption{\imk{Survival probability of a single-site initial wavepacket in the NEE phase for $\gamma=2$ and fractal disorder with (a)~$d= 0.6$ and (b)~$d=0.8$ vs time $t$ rescaled by $\Gamma$. The raw data with no rescaling of $t$ is plotted in (c) and (d) showing clear separation between different $L$}.}
\label{fig:Rprob}
\end{figure}
In Fig.~\ref{fig:Rprob} we see that \imk{the rescaling of time by the analytical expression $\Gamma$ collapses the curves for different $L$ up to small values of $\mathcal{R}(t)$ (clearly there is no collapse if the time is not rescaled). Discrepancy at later times might be given by non-self-averaging nature of the fractal spectrum.}
The exponential behavior of the return probability is given by the Lorenzian wave-function profile, Eq.~(9), in the energy domain. Similarly to the standard RP case~\cite{RP_R(t)_2018}, one can calculate the return probability from its Fourier transform $K(\omega)$, which \imk{also takes the Lorenzian form}. \imk{Here $\Gamma$ plays the role of the energy/frequency scale, below which levels are Wigner-Dyson correlated, while otherwise their statistics is given by the one of the diagonal elements (fractal in this case).}
\bibliography{Lib}